\documentclass[
prd,
twocolumn,%!TEX encoding = UTF-8 Unicode
nofootinbib,
showpacs ,amssymb, aps]{revtex4}
\usepackage{amsmath}
\usepackage{float}

\usepackage{color}

\usepackage{ulem}
\usepackage{cancel}

\usepackage{comment}
\usepackage{graphicx}

\usepackage{amsmath,amssymb}
\usepackage{bm}

\usepackage{natbib}
\usepackage{wasysym}

\newcommand{\I}{{\mathcal I}}
\newcommand{\R}{{\mathcal R}}
\newcommand{\A}{{\mathcal A}}
\newcommand{\sn}{\mbox{sn}}
\newcommand{\cn}{\mbox{cn}}
\newcommand{\dn}{\mbox{dn}}

\begin{document}

\title{
	Tilted torus magnetic fields in neutron stars and their gravitational wave signatures
}

\author{Paul D. Lasky}
	\email{paul.lasky@unimelb.edu.au}
\author{Andrew Melatos}
	\email{amelatos@unimelb.edu.au}
	\affiliation{School of Physics, University of Melbourne, Parkville VIC 3010, Australia}

		\begin{abstract}
			Gravitational-wave diagnostics are developed for discriminating between varieties of mixed poloidal-toroidal magnetic fields in neutron stars, with particular emphasis on differentially rotating protoneutron stars.  It is shown that {\it tilted torus} magnetic fields, defined as the sum of an internal/external poloidal component, whose axis of symmetry is tilted with respect to the rotation axis, and an internal toroidal component, whose axis of symmetry is aligned with the rotation axis, deform the star triaxially, unlike twisted torus fields, which deform the star biaxially.  Utilizing an analytic tilted torus example, we show that these two topologies can be distinguished by their gravitational wave spectrum and polarization phase portraits.  For example, the relative amplitudes and frequencies of the spectral peaks allows one to infer the relative strengths of the toroidal and poloidal components of the field, and the magnetic inclination angle.  Finally, we show how a tilted torus field arises naturally from magnetohydrodynamic simulations of differentially rotating neutron stars, and how the gravitational wave spectrum evolves as the internal toroidal field winds up.  These results point to the sorts of experiments that may become possible once gravitational wave interferometers detect core-collapse supernovae routinely.
		\end{abstract}
		
	%	\received{}
	%	\revised{}
	%	\accepted{}
		
		%\ccc{}
		%\cpright{}{}
		
		%\keywords{Neutron Stars, Magnetars, Magnetic Field Equilibria}

		\maketitle

\section{Introduction}
Strongly magnetized neutron stars are candidates for detection by ground-based, long-baseline, interferometric gravitational wave detectors such as the Laser Interferometer Gravitational-Wave Observatory (LIGO) and Virgo \cite{abbott09a}.  For example, in magnetars born with millisecond spin periods, strong differential rotation coupled with turbulent convection drives an $\alpha$-$\Omega$ dynamo, which winds the internal magnetic field as high as $\sim10^{16}\,\mbox{G}$ \cite{duncan92,thompson93}.  Such strong fields deform the star sufficiently to emit gravitational waves \cite[e.g.,][]{ioka01,palomba01} detectable out to Virgo cluster distances \cite{stella05,dallosso09}.  Indeed, a magnetar containing a strong toroidal field evolves on the viscous dissipation time-scale to become an orthogonal rotator, maximising its gravitational wave luminosity \cite{cutler02}.    

Even ordinary pulsars, with surface dipole magnetic fields of order $10^{12}\,\mbox{G}$, can emit detectable gravitational-wave signals, if the interior exists in an exotic thermodynamic phase like a color superconductor, which increases the ellipticity $10^{3}$-fold \cite{glampedakis12}.  Non-detection from the Crab limits its internal magnetic field to $\lesssim10^{16}\,\mbox{G}$ $(10^{13}\,\mbox{G}$) if the core is (not) a color superconductor  \cite{glampedakis12,abbott08c}.

%Young pulsars like Vela and the Crab may also be good sources for gravitational wave observations.  Non-detection from the Crab limits its internal magnetic field to $\lesssim10^{16}\,\mbox{G}$ \cite{abbott08c}, more than three-orders of magnitude larger than the inferred surface dipole component.  This result assumes the matter is in a normal fluid state.  If it is a color superconductor, the stellar ellipticity increases $10^{3}$-fold \cite{glampedakis12} and gravitational waves from Crab-like pulsars are detectable if the ratio of internal to surface field is $\gtrsim10$.

Many authors have studied how the magnetic field {\it topology} affects the gravitational wave signal.  Early work concentrated on purely poloidal \cite{bonazzola96,ioka01,palomba01} or purely toroidal \cite{cutler02} fields, although these are dynamically unstable \cite{kruskal54,tayler57,tayler73,wright73,markey73,markey74}.  Recently, large-scale, non-linear numerical simulations have investigated more realistic magnetic field configurations, e.g.\cite{geppert06,braithwaite06a,braithwaite06b,braithwaite06,braithwaite07,braithwaite08,braithwaite09,kiuchi11,lasky11,ciolfi12,lasky12}.  Chief among these is the `twisted torus', defined as a poloidal component of low multipole order, which exists inside and outside the star, and a toroidal component that threads the closed-field-line region of the internal poloidal field.  Twisted tori create axisymmetric deformations.  They have been used to calculate the gravitational wave signal from barotropic \cite{haskell08} and non-barotropic \cite{mastrano11} stars and in general relativity \cite{ciolfi10}.  Despite the popularity of twisted tori, their stability remains an open question \cite{lander12,akgun13,ciolfi13}.

In this article we investigate an alternative magnetic configuration: the `tilted torus', defined as the sum of an internal/external poloidal component, whose axis of symmetry is tilted with respect to the rotation axis, and an internal toroidal component, whose axis of symmetry is aligned with the rotation axis\footnote{\label{footnoteref}Strictly speaking, the tilted `poloidal' field has a non-zero component in the azimuthal direction.  Likewise, in the tilted frame, the `toroidal' field is no longer purely toroidal.  To keep the terminology simple, we persist in referring to these components as poloidal and toroidal, even when they are misaligned.}.  A picture of a tilted torus is presented in figure \ref{tilted} (see section \ref{model} for details).  Tilted tori are motivated by conditions inside a protoneutron star, where differential rotation \cite[e.g.,][]{duncan92,thompson93,wheeler00,wheeler02} or $r$-mode instabilities \cite{rezzolla00,rezzolla01a,rezzolla01b,cuofano12a} wind up the internal field.  If the progenitor's field is tilted with respect to the rotation axis, the resulting magnetic configuration contains two misaligned components.  In section \ref{numerical} of this article, we demonstrate qualitatively how a tilted torus arises naturally in this way from a magnetohydrodynamic (MHD) simulation of a differentially rotating protoneutron star with an inclined poloidal field.  Other simulations have shown that the resulting transient may be transitory or not \cite{braithwaite08,braithwaite09,lasky11,ciolfi11,kiuchi11,lasky12,ciolfi12}.

We emphasize that the tilted torus is a physically motivated toy model; it is not a substitute for systematic numerical studies, which are outside the scope of this paper.  Nevertheless, the toy model plays a valuable role in revealing what practical things can be learned from upcoming gravitational wave observations, especially at the modest signal-to-noise expected for the first detections.  In particular, the toy model suffices to demonstrate the central result of the paper: that a tilted torus creates a non-axisymmetric stellar deformation, unlike a twisted torus, adding new lines to the gravitational wave spectrum and allowing an observer to distinguish between the two topologies in principle.  In practice, realistic field configurations, both axisymmetric and otherwise, are likely to be more complicated than twisted and tilted tori respectively.  Regardless how complicated though, any magnetic field configuration leads to a moment-of-inertia tensor with three eigenvalues, implying that realistic magnetic field structures cannot be inferred uniquely from a typical set of gravitational wave observations.  The tilted torus is an example of a toy model of a differential-rotation-dominated field which can be distinguished from a twisted torus using gravitational wave measurements, at least in principle, even though its parameters (e.g., tilt angle) cannot be inferred uniquely.  Here we take the first step towards analysing the spectrum and understanding exactly what can, and cannot, be inferred from it for a given level of signal to noise.  

The article is set out as follows.  In section \ref{magfield} we construct a representative tilted torus, solving the MHD force-balance equation to derive the density and pressure perturbations in \ref{perturb} and the mass quadrupole moment in \ref{massquad}.  Applying the formulae in Ref. \cite{zimmerman80}, we calculate the gravitational wave signal from a biaxial star (i.e., twisted torus) in \ref{biaxial} and a triaxial star (i.e., tilted torus) in \ref{smallwobble} (in the small wobble approximation) and \ref{arbitrarywobble} (arbitrary wobble).  We introduce a new diagnostic tool to assist with this task: the phase portrait in the $h_{+}$-$h_{\times}$ plane, where $h_{+}$ and $h_{\times}$ are the gravitational wave strains in the plus and cross polarizations respectively.  In section \ref{numerical} we apply the results to the output from the three-dimensional, general relativistic MHD solver {\sc horizon} \cite{zink11,lasky11,zink12,lasky12}, motivating further the tilted torus by building a similar magnetic configuration in a differentially rotating neutron star with an initially poloidal field and showing how the gravitational wave spectrum evolves as the magnetic field winds up.  We conclude in section \ref{conclusion} by detailing a recipe for how the internal magnetic field geometry can be inferred from future gravitational wave observations.

\section{Hydromagnetic Equilibrium}\label{model}
We treat the magnetic field as a perturbation on a spherically symmetric star.  The force-balance equation can be expressed to first order in magnetic pressure as
\begin{align}
	\mu_{0}^{-1}\left(\nabla\times{\bf B}\right)\times{\bf B}=\nabla\delta p+\delta\rho\nabla\Phi.\label{forcebalance}
\end{align}
Here, $\Phi(r)$ is the background Newtonian potential, $\delta\rho$ and $\delta p$ are the density and pressure perturbations respectively, and we work in the Cowling approximation, $\delta\Phi=0$.  Omitting gravitational perturbations affects ellipticity calculations by up to a factor two \cite{yoshida13}\footnote{The Cowling approximation neglects corrections of order $\delta\rho/\rho\propto\epsilon\propto B^{2}$.  \citet{yoshida13} showed that the corrections approach a factor of two when the surface magnetic field strength is $B\approx4.4\times10^{16}\,{\rm G}$, at the upper limit of magnetic field strengths expected in protomagnetars.}, which is tolerable; the purpose of this article is not to make precise predictions for gravitational-wave amplitudes but to provide a phenomenological framework for interpreting gravitational wave spectra from magnetically triaxial stars.  Equation (\ref{forcebalance}) is solved in Ref. \cite{mastrano11} for an axisymmetric magnetic field and a non-barotropic equation of state.  Here we generalize \cite{mastrano11} by rotating the poloidal component of the field, misaligning it with the toroidal symmetry axis.  Spherical polar coordinates are used throughout, with $r$ expressed in units of the stellar radius.

\subsection{Tilted torus}\label{magfield}
The magnetic field is described using two coordinate systems, ${\bf x}$ and ${\bf x}'$, where the primed coordinates are rotated by an angle $\alpha$ in the $x$-$z$ plane with respect to the unprimed frame.  The total field can be expressed as the sum of a `poloidal' and `toroidal' component (see footnote \ref{footnoteref}),
\begin{align}
	{\bf B}({\bf x})={\bf B}_{p}({\bf x})+{\bf B}_{t}({\bf x}).\label{B}
\end{align}
The poloidal field component, ${\bf B}_{p}$, is axially symmetric around the $z'$-axis, matches continuously to an external dipole field, and is sourced by a finite and continuous current density everywhere in the star \cite{mastrano11}.  The toroidal component, ${\bf B}_{t}$, is symmetric about the rotation axis (i.e., the $z$-axis), so that $\alpha$ is the angle between the rotation axis and the poloidal component's axis of symmetry.  

The poloidal field is given the same functional form as Ref.  \cite{mastrano11} in the primed coordinates:
\begin{align}
	{\bf B}_{p}=B_{0}\eta_{p}\nabla'\gamma\left(r,\theta'\right)\times\nabla'\phi'.\label{Bp}
\end{align}
Here, $\eta_{p}$ is a dimensionless parameter defining the relative strength of the poloidal and toroidal components, $\nabla'$ is the gradient operator in the primed coordinates and $\gamma\left(r,\theta'\right)=f(r)\sin^{2}\theta'$ is a flux function.  The radial function, $f(r)$, enjoys considerable freedom (for details see \cite{mastrano11,akgun13,mastrano13}).  We choose
\begin{align}
	f(r) = \frac{35}{8}\left(r^{2}-\frac{6r^{4}}{5}+\frac{3r^{6}}{7}\right),\label{fr}
\end{align}
ensuring that the magnetic field is continuous with a pure dipole outside the star, the current density is finite at the origin, and there are no surface currents.

The toroidal field is defined following Ref. \cite{mastrano11}, 
\begin{align}
	{\bf B}_{t}=B_{0}\eta_{t}\beta\left[\gamma\left(r,\theta\right)\right]\nabla\phi.\label{Bt}
\end{align}
An integrability condition for equation (\ref{forcebalance}) follows from $\partial_{\theta}\partial_{\phi}\delta p=\partial_{\phi}\partial_{\theta}\delta p$, which constrains $\beta(\gamma)$.  Two classes of solutions to the integrability condition exist.  The simplest non-trivial class has
\begin{align}	
	\beta\left(\gamma\right)=
	\left\{
		\begin{array}{lcl}
		\gamma-1 & {\rm for} & \gamma\ge1\\
		0 & {\rm for} & 0<\gamma<1
		\end{array}
	\right..\label{beta}
\end{align}
It is worth noting that such a toroidal field distribution leads to non-zero surface currents.  A better model is ultimately needed (see sections \ref{numerical} and \ref{conclusion}), but surface currents do not interfere with the goal of using idealized field configurations to illustrate how to infer the interior field topology from gravitational wave signals.

\begin{figure*}
	\begin{center}
	\includegraphics[angle=0,width=0.95\textwidth]{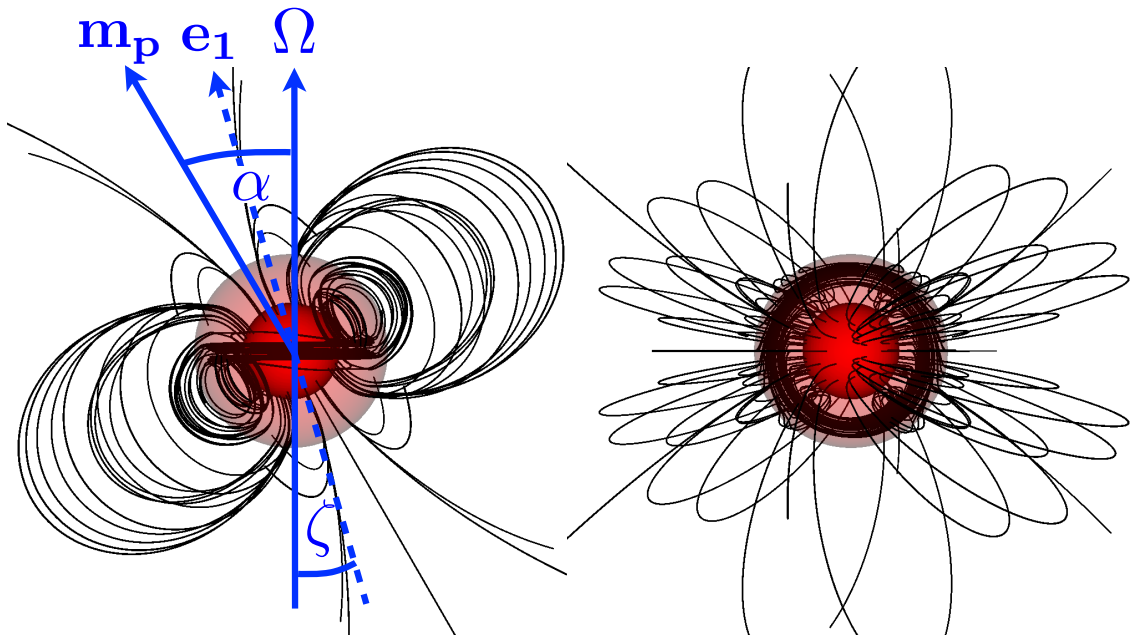}
	\end{center}
	\caption{\label{tilted}Magnetic field lines for a tilted torus.  The poloidal axis of symmetry, ${\bf m}_{p}$, makes an angle $\alpha$ with the rotation axis, ${\bf\Omega}$, which is also the axis of symmetry of the toroidal component.  The principal axis of inertia, ${\bf e}_{1}$, makes an angle $\zeta$ with ${\bf\Omega}$.  The figure has $\eta_{t}=10^{3}\eta_{p}$ to artificially emphasize the toroidal component of the field.  The semi-transparent red and opaque red contours are at half the star's radius and the star's radius respectively.  The left panel shows the star from the equator, and the right panel looks down the rotation axis.  One can see two groups of toroidal field lines: the toroidal component of ${\bf B}_{p}$, which appears as the flower-like pattern in the outer parts of the figure, and ${\bf B}_{t}$, which appears as the densely packed annulus of circular curves in the plane of the page. 
	}
\end{figure*}

A three-dimensional plot of the magnetic field lines is presented in figure \ref{tilted} with $\eta_{t}/\eta_{p}=10^{3}$ to emphasize the toroidal field component.  The poloidal axis of symmetry, ${\bf m}_{p}$, is tilted with respect to the rotation axis, ${\bf\Omega}$, by the angle $\alpha$.  We show below that the principal axis of inertia, ${\bf e}_{1}$, lies somewhere between ${\bf m}_{p}$ and ${\bf\Omega}$.  The angle between ${\bf e}_{1}$ and ${\bf\Omega}$ is denoted by $\zeta$.

The relative strength of the poloidal component is measured through the ratio of the poloidal to the total magnetic energy within the star,
\begin{align}
	\Lambda=
	%\frac{E_{\rm mag}^{\rm pol}}{E_{\rm mag}^{\rm tot}}=
	\frac{\eta_{p}^{2}}{\eta_{p}^{2}+q\eta_{t}^{2}}.\label{Lambda}
\end{align}
For the magnetic field configuration described by (\ref{B})--(\ref{beta}), we find $q=1.95\times10^{-4}$.

\subsection{Density and pressure perturbations}\label{perturb}
The $\theta$ and $\phi$ components of the force balance equation respectively contain $\partial_{\theta}\delta p$ and $\partial_{\phi}\delta p$ terms.  Integrating the $\phi$ component with respect to $\phi$ gives an arbitrary function of $r$ and $\theta$.  Substituting into the $\theta$ component, one can solve for $\delta p$ up to an arbitrary function of $r$, which does not affect the quadrupole moment and hence the gravitational wave signal.  After some algebra, we obtain
\begin{widetext}
\begin{align}
	\frac{\mu_{0}\delta p}{B_{0}^{2}}=&\frac{\eta_{p}^{2}f}{r^{2}}\left(\frac{d^{2}f}{dr^{2}}-\frac{2f}{r^{2}}\right)\left(\sin\alpha\sin\theta\cos\phi-\cos\alpha\cos\theta\right)^{2}
		-\frac{2\eta_{p}\eta_{t}f}{r^{2}}\frac{df}{dr}\sin\alpha\sin\theta\sin\phi
		+\frac{\eta_{t}^{2}f}{r^{2}}\left[1-\gamma\left(r,\theta\right)+\ln\gamma\left(r,\theta\right)\right],\label{dp}
\end{align}
Equation (\ref{dp}) applies for $\gamma(r,\theta)\ge1$; for $\gamma(r,\theta)<1$ and hence $\eta_{t}=0$, only the first term on the right-hand side survives.  Substituting $\delta p$ into the radial component of (\ref{forcebalance}), we arrive at\begin{align}
	\frac{\mu_{0}M(r)G\delta\rho}{B_{0}^{2}}=&-\eta_{p}^{2}fr^{2}\left\{\frac{d}{dr}\left[\frac{1}{r^{2}}\left(\frac{d^{2}f}{dr^{2}}-\frac{2f}{r^{2}}\right)\right]\left(\sin\alpha\sin\theta\cos\phi-\cos\alpha\cos\theta\right)^{2}+\frac{1}{fr^{2}}\frac{df}{dr}\left(\frac{d^{2}f}{dr^{2}}-\frac{2f}{r^{2}}\right)\right\}\notag\\
		&+\eta_{p}\eta_{t}\frac{\sin\phi\sin\alpha}{\sin\theta}\left[\left(\frac{d^{2}f}{dr^{2}}-\frac{2f}{r^{2}}\right)\left(1+f\sin^{2}\theta\right)+\sin^{2}\theta\frac{df}{dr}\left(\frac{df}{dr}-\frac{2f}{r}\right)^{2}\right]\notag\\
		&-\eta_{t}^{2}\left(\frac{df}{dr}-\frac{2f}{r}\right)\left[1-f\sin^{2}\theta+\ln\left(f\sin^{2}\theta\right)\right],\label{drho}
\end{align}
\end{widetext}
where $M(r)=4\pi\int_{0}^{r'}dr'r'^{2}\rho(r')$ is the background mass function defined in terms of the unperturbed density, for which we adopt an idealized form,\begin{align}
	\rho(r)=\rho_{c}\left(1-r^{2}\right),\label{rhob}
\end{align}
where $\rho_{c}$ is the central density.  This choice of density profile was used in \cite{mastrano11}, where ellipticity calculations were shown to be accurate to a few percent when compared with an $n=1$ polytrope.  We refer the reader to Refs. \cite{mastrano11,akgun13} for further justification of this choice.  Throughout the article we use a $1.4\,M_{\odot}$ background star.

\subsection{Mass quadrupole moment}\label{massquad}
Equations (\ref{drho}) and (\ref{rhob}) can be used to calculate the moment-of-inertia tensor,
\begin{align}
	\I^{{\rm tot}}_{ij}&=\int_{V}\left[\rho(r)+\delta\rho(r,\theta,\phi)\right]\left(r^{2}\delta_{ij}-x_{i}x_{j}\right)dV.
\end{align}
The eigenvalues and eigenvectors of $\I_{ij}$ correspond to the principal moments and axes, which govern the gravitational wave signal from a triaxial neutron star \cite{zimmerman79,zimmerman80}.  
%The axis ${\bf e}_{1}$ in figure \ref{tilted} corresponds to the largest eigenvalue.  

\begin{figure}
	\begin{center}
	\includegraphics[angle=0,width=0.95\columnwidth]{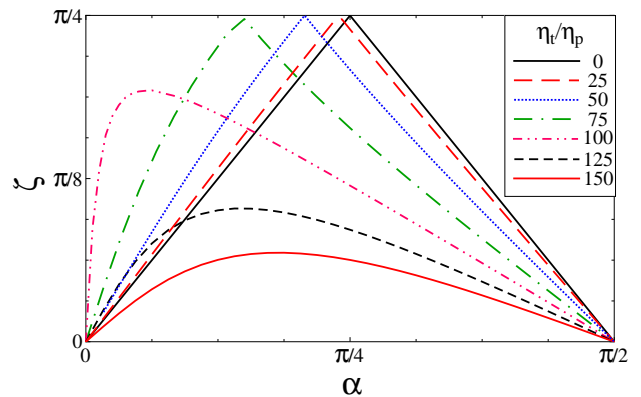}
	\end{center}
	\caption{\label{angles}Wobble angle $\zeta$ between the star's principal axis and rotation axis as a function of the magnetic inclination $\alpha$, for different ratios of the toroidal to poloidal magnetic field strengths, $\eta_{t}/\eta_{p}$.}
\end{figure}

How do the principal axes of inertia relate to the tilted torus geometry in figure \ref{tilted}?  To get a feel for this, we plot the wobble angle $\zeta$ as a function of $\alpha$ in figure \ref{angles}.  Each curve on this figure represents a contour of constant $\eta_{t}/\eta_{p}$.  Clearly, when the two components are aligned (i.e., $\alpha=0$), ${\bf e}_{1}$ is also aligned with the symmetry axis of the fields, i.e. $\zeta=0$.  The same is true when the two components are perpendicular; the poloidal component has one principal eigenvector oriented orthogonal to the $z$-axis and another parallel to the $z$-axis.  In between, an intermediate strength toroidal field (i.e., $0<\eta_{t}/\eta_{p}\lesssim100$) causes $\zeta>\alpha$ for $\alpha\lesssim\pi/4$, whereas a stronger toroidal component (i.e., $\eta_{t}/\eta_{p}\gtrsim100$) implies the wobble angle is closer to the rotation axis for $\alpha\lesssim\pi/4$.

\section{Gravitational Wave Signal}\label{GWs}
The gravitational waveform for a triaxial, precessing, rigid body was first written down by \citet{zimmerman80}, following earlier work in the small-wobble-angle limit \cite{zimmerman79}.  The gravitational wave amplitude depends on the principal moments of inertia\footnote{We assume $J^{2}>2E\I_{2}$, where $J=(\sum_{i}\I_{i}^{2}\Omega_{i}^{2})^{1/2}$ and $E=(\sum_{i}\I_{i}\Omega_{i}^{2})/2$ are respectively the angular momentum and rotational energy; otherwise one interchanges $\I_{1}$ and $\I_{3}$ \cite{zimmerman80}.}, $\I_{1}<\I_{2}<\I_{3}$, the initial values of the components of the body's angular velocity, $a=\Omega_{1}(t=0)$ and $b=\Omega_{3}(t=0)$ (along the ${\bf e}_{1}$ and ${\bf e}_{3}$ axes respectively), and the inclination angle, $\iota$.  We define two ellipticities according to 
\begin{align}
	e_{1}&=\left[2\left(\I_{3}-\I_{1}\right)/\I_{1}\right]^{1/2},\label{e1}\\
	e_{2}&=\left[2\left(\I_{3}-\I_{2}\right)/\I_{2}\right]^{1/2},\label{e2}
\end{align}
and a mean ellipticity as
\begin{align}
	\epsilon=e_{1}e_{2}/2.\label{ecc}
\end{align}

The $h_{+}$ and $h_{\times}$ gravitational waveforms are respectively
\begin{align}
	h_{+}=&d^{-1}\big[\left(\R_{yj}\cos\iota-\R_{zj}\sin\iota\right)\left(\R_{yk}\cos\iota-\R_{zk}\sin\iota\right)\notag\\
	&-\R_{xj}\R_{xk}\big]\A_{jk},\label{hplus}\\
	h_{\times}=&2d^{-1}\left(\R_{yj}\cos\iota-\R_{zj}\sin\iota\right)\R_{xk}\A_{jk}.\label{hcross}
\end{align}
Here, $d$ is the distance to the source, $\A_{ij}$ is given by
\begin{align}
	\A_{jk}=&-2|{\mathbf\Omega}|^{2}\I_{jk}+\left(\epsilon_{\ell mj}\dot{\Omega}_{\ell}+\Omega_{m}\Omega_{j}\right)\I_{mk}\notag\\
			&+\left(\epsilon_{\ell mk}\dot{\Omega}_{\ell}+\Omega_{m}\Omega_{k}\right)\I_{jm}+2\epsilon_{\ell mj}\epsilon_{npk}\Omega_{\ell}\Omega_{n}\I_{mp},\label{Ajk}
\end{align}
and $\R_{\mu j}$ is the rotation matrix in terms of the Euler angles, $\theta$, $\varphi$ and $\psi$,
\begin{widetext}
\begin{align}
	\R=\left(\begin{array}{ccc}
		\cos\psi\cos\varphi-\cos\theta\sin\psi\sin\varphi & -\sin\psi\cos\varphi-\cos\theta\cos\psi\sin\varphi & \sin\theta\sin\varphi\\
		\cos\psi\sin\varphi+\cos\theta\sin\psi\cos\varphi & -\sin\psi\sin\varphi+\cos\theta\cos\psi\cos\varphi & -\sin\theta\cos\varphi\\
		\sin\theta\sin\psi & \sin\theta\cos\psi & \cos\theta
		\end{array}\right).
\end{align}
\end{widetext}

The individual components of the angular velocity vector, $\Omega_{i}$, are periodic in time, with period $T$ defined by
\begin{align}
	T=\frac{4K(m)}{b}\left[\frac{\I_{1}\I_{2}}{\left(\I_{3}-\I_{2}\right)\left(\I_{3}-\I_{1}\right)}\right]^{1/2},\label{T}
\end{align}
where $K(m)$ is the complete elliptic integral of the first kind, with
\begin{align}
	m=\frac{\left(\I_{2}-\I_{1}\right)\I_{1}a^{2}}{\left(\I_{2}-\I_{3}\right)\I_{2}b^{2}}.
\end{align}
If the oblateness is small (i.e. $\I_{1}\approx\I_{2}$), then the timescale $T$ becomes long, and one recovers the axisymmetric results, with $T\rightarrow2\pi\I_{1}/\left[\Omega_{3}\left(\I_{3}-\I_{1}\right)\right]$ being the usual free-precession period.  We discuss this limit in more detail at the end of the current section.  The body-frame angular velocity components are then expressed in terms of the temporal variable $\tau=4K(m)t/T$ as
\begin{align}
	\Omega_{1}=&a\,\cn\left(\tau,\,m\right),\\
	\Omega_{2}=&a\left[\frac{\I_{1}\left(\I_{3}-\I_{1}\right)}{\I_{2}\left(\I_{3}-\I_{2}\right)}\right]^{1/2}\sn\left(\tau,\,m\right),\\
	\Omega_{3}=&b\,\dn\left(\tau,\,m\right),
\end{align}
where $\sn$, $\cn$ and $\dn$ are the Jacobi elliptic functions.

The $\theta$ and $\psi$ Euler angles have period $T/2$, and are expressed explicitly as
\begin{align}
	\cos\theta=&\frac{\I_{3}b}{J}\dn(\tau,m),\\
	\tan\psi=&\left[\frac{\I_{1}\left(\I_{3}-\I_{2}\right)}{\I_{2}\left(\I_{3}-\I_{1}\right)}\right]^{1/2}\frac{\cn(\tau,m)}{\sn(\tau,m)}.
\end{align}
The Euler angle, $\varphi$, is expressed as $\varphi=\varphi_{1}+\varphi_{2}$, with 
\begin{align}
	\exp\left[2i\varphi_{1}(t)\right]=\frac{\vartheta_{4}\left(2\pi t/T-i\pi A,\,q\right)}{\vartheta_{4}\left(2\pi t/T+i\pi A,\,q\right)}
\end{align}
where $\vartheta_{4}(u,\,q)$ is the fourth elliptic theta function with nome $q=\exp[-\pi K(1-m)/K(m)]$, and $A$ is a solution of $\sn(2iA K)=i\I_{3}b/(\I_{1}a)$.  The second component is linear in time
\begin{align}
	\varphi_{2}(t)=\frac{2\pi}{T'} t,\label{varphi2}
\end{align}
with
\begin{align}
	\frac{2\pi}{T'}=\frac{J}{\I_{1}}-\frac{2i}{T}\frac{\vartheta_{4}'\left(i\pi A,\,q\right)}{\vartheta_{4}\left(i\pi A,\,q\right)}\label{2pitprime}
\end{align}
where $\vartheta_{4}'(u,\,q)$ is the derivative of $\vartheta_{4}$ with respect to $u$.  

The highest spectral peak for the triaxial models shown in section \ref{triaxial} has period $T'$, as in Ref. \cite{zimmerman80}.  In the biaxial limit (i.e. $\I_{1}\rightarrow\I_{2}$), one finds $q=0$, $\vartheta_{4}(u,0)=1$, $\vartheta_{4}'(u,0)=0$ and hence $T'=2\pi\I_{1}/J$, which is the period of the largest spectral peak for biaxial stars (see section \ref{biaxial})\cite{zimmerman80}.  We discuss this limit further in the following sections.

%The above set of equations are used below to calculate the gravitational wave emission from a triaxial neutron star modelled as set out in section \ref{model}.

\section{Poloidal Field}\label{biaxial}
In our model, stars with $\eta_{t}=0$ are necessarily biaxial.  One principal axis coincides with the symmetry axis of the poloidal magnetic field, implying that the star precesses for all $\alpha\neq0$.  The gravitational wave emission of such systems is well known \cite{bonazzola96}.  We examine it briefly here for three reasons: (i) to double-check the formulas in section \ref{GWs} against known results; (ii) to establish a baseline against which to interpret the general, triaxial case in sections \ref{triaxial} and \ref{numerical}; and (iii) to introduce a new diagnostic tool, the polarization phase portrait, defined as the parametric curve $[h_{+}(t),\,h_{\times}(t)]$ in the $h_{+}$-$h_{\times}$ plane.  We consider an artificially strong poloidal field of $10^{17}\,\mbox{G}$ to decrease the precession period, so that it is clearly visible by eye in the Fourier transforms and phase portraits presented below.  
%Our recipe applies most readily to newly born neutron stars, with magnetic fields on the order of $10^{14}$--$10^{16}\,\mbox{G}$\cite{duncan92,thompson93}, but it is also potentially applicable to radio pulsars with weaker fields, a point we discuss further below.  

\begin{figure*}
	\begin{center}
	\includegraphics[angle=0,width=.95\textwidth]{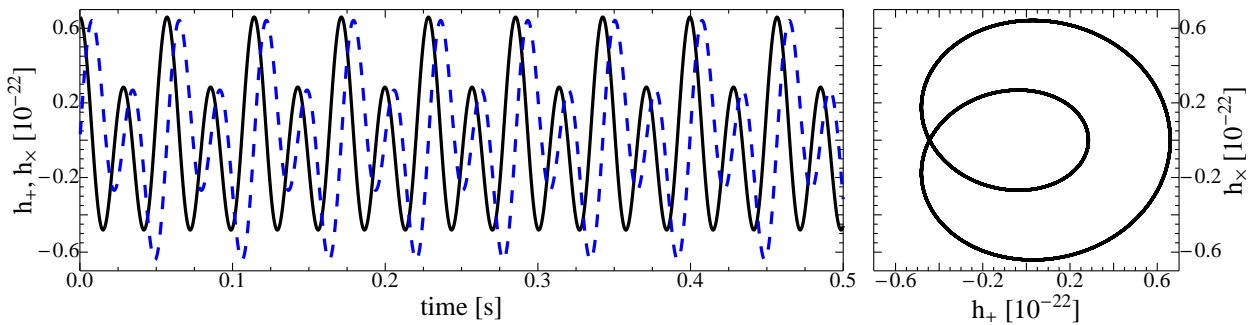}
	\end{center}
	\caption{\label{htime1}Gravitational wave signal from a biaxial star deformed by a purely poloidal magnetic field.  Left panel:  Wave strain in the plus polarisation $h_{+}$ (solid black curve) and cross polarization $h_{\times}$ (dashed blue curve) as functions of time for $\eta_{t}=0$ ($\Lambda=1$), $B_{0}=10^{17}\,\mbox{G}$, $\alpha=\iota=\pi/4$, $\Omega_{0}/2\pi=100\,\mbox{Hz}$ and $d=1\,\mbox{kpc}$.  Right panel: Polarization phase portrait in the $h_{+}$-$h_{\times}$ plane.}
\end{figure*}

In the left panel of figure \ref{htime1} we plot $h_{+}$ (solid black curve) and $h_{\times}$ (dashed blue curve) as functions of time for a purely poloidal example with $\alpha=\iota=\pi/4$ and $\Omega_{0}/2\pi=100\,\mbox{Hz}$.  The system emits at angular frequencies $\omega$ and $2\omega$, with $\omega=J/I_{1}=1.1007\Omega_{0}$.  In the right panel of figure \ref{htime1} we plot the polarization phase portrait.  The curve closes because the emission frequencies, $\omega$ and $2\omega$, are commensurable.  

\begin{figure*}
	\begin{center}
	\includegraphics[angle=0,width=.95\textwidth]{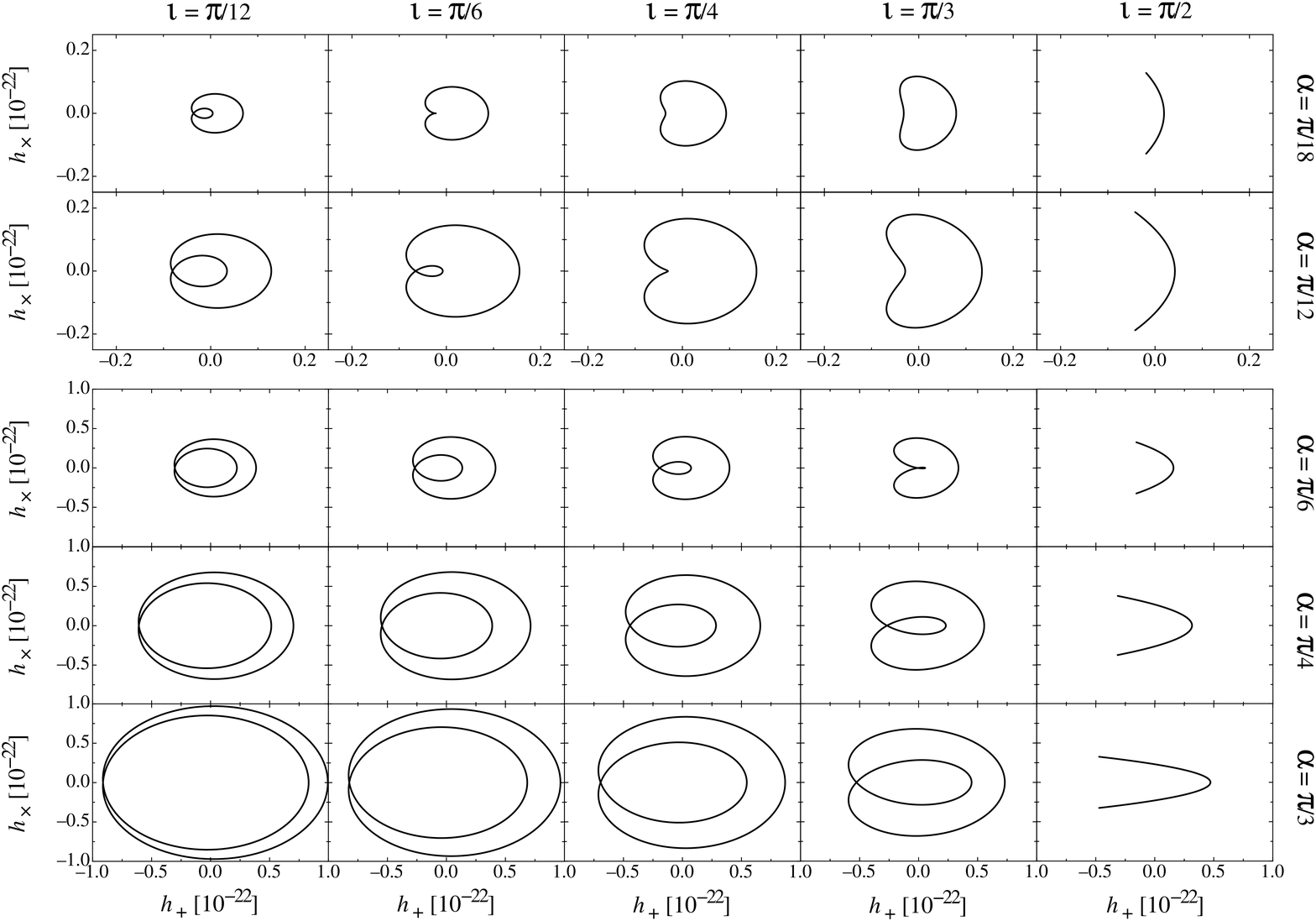}
	\end{center}
	\caption{\label{PhaseBiaxial}Polarisation phase portraits, $[h_{+}(t),\,h_{\times}(t)]$, for biaxial stars (i.e. zero toroidal field, $\eta_{t}=0$) covering a range of magnetic inclination angles ($\pi/18\le\alpha\le\pi/2$; top to bottom rows) and observer inclination angles ($\pi/12\le\iota\le\pi/2$; left to right columns).  Note the change of scale between the top two rows and the bottom three rows.  The stellar parameters are $B_{0}=10^{17}\,\mbox{G}$ and $\Omega_{0}/2\pi=100\,\mbox{Hz}$.}
\end{figure*}

Figure \ref{PhaseBiaxial} displays a grid of phase portraits for a biaxial star with $B_{0}=10^{17}\,\mbox{G}$, $\Omega_{0}/2\pi=100\,\mbox{Hz}$, and 25 different combinations of ($\alpha$, $\iota$).  The cases $\alpha=0$ (zero emission) and $\alpha=\pi/2$ (emission at a single frequency; phase portrait is an ellipse) are omitted.  The phase portraits highlight the relative amplitudes of the two spectral components.  For example, in the bottom left panel ($\alpha=\pi/3$, $\iota=\pi/12$), the components have similar amplitudes; the inner and outer ovals nearly overlap.  Importantly, the curves in every panel close as the star is biaxial; this includes the $\iota=\pi/2$ portraits which are degenerate (the curve traces back and forth along the `U' shape).  

\section{Tilted torus}\label{triaxial}
Neutron stars with misaligned poloidal and toroidal field components ($\eta_{t}/\eta_{p}\neq0$ and $\alpha\neq0$) are necessarily triaxial.  These tilted torus field configurations are idealized models which seek to represent qualitatively some of the generic features (e.g., differential rotation and tilted magnetic field axis) that may be present, perhaps as transients, in newly born neutron stars, where differential rotation or $r$-mode instabilities wind up the internal toroidal component around the rotation axis, and the misaligned poloidal component is a fossil of the protoneutron star's field.  The wobble angle depends on both $\eta_{t}/\eta_{p}$ and $\alpha$ (see figure \ref{angles}).  We first explore the gravitational wave signal from a triaxial star in the small wobble angle limit, and subsequently analyse the problem in full generality.  

\subsection{Small wobble angle}\label{smallwobble}
Equations (\ref{hplus})--(\ref{2pitprime}) can be approximated to first \cite{zimmerman79,zimmerman80} and second order \cite{vandenbroeck05} in the wobble angle, $\zeta$.
%, when the latter is small.  
At second order, gravitational waves are emitted at four frequencies, which are, in increasing order,
\begin{align}
	\frac{\left(1-\epsilon\right)\omega}{1+\epsilon},\,\,\,
	\omega,\,\,\,
	\frac{2\omega}{1+\epsilon},\,\,\rm{and}\,\,\,
	2\omega.
\end{align}

\begin{figure*}
	\begin{center}
	\includegraphics[angle=0,width=.95\textwidth]{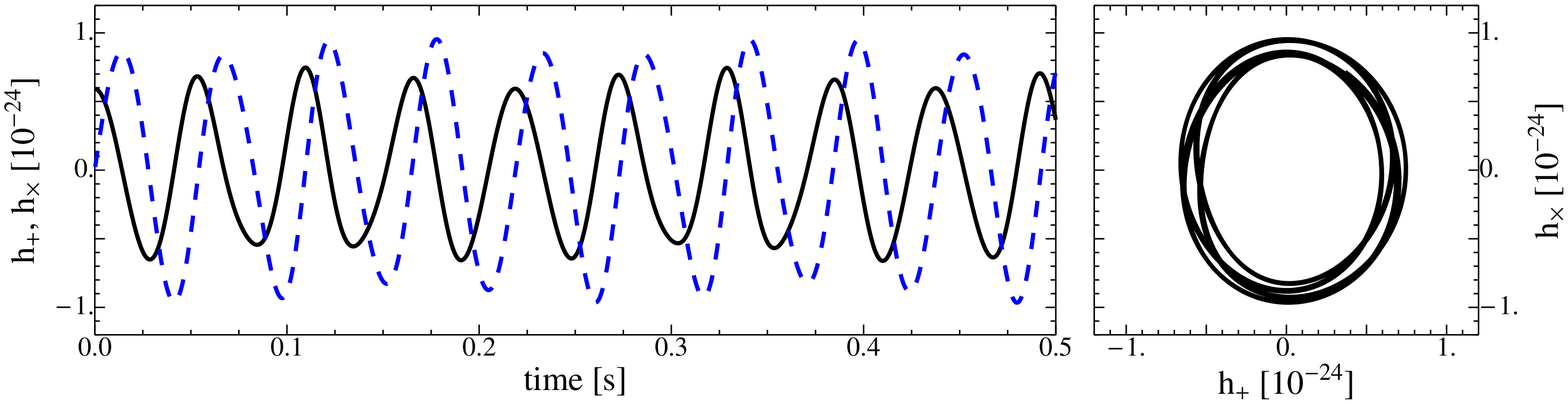}
	\end{center}
	\caption{\label{htime1b}Gravitational wave signal from a triaxial star deformed by a tilted torus magnetic field.  Left panel: Wave strain in the plus polarization $h_{+}$ (solid black curve) and cross polarization $h_{\times}$ (dashed blue curve) as functions of time for $\eta_{t}/\eta_{p}=50$, $B_{0}=10^{17}\,\mbox{G}$, $\iota=\pi/4$, $\alpha=\pi/180$ and $d=1\,\mbox{kpc}$.  Right panel:  Polarization phase portrait in the $h_{+}$-$h_{\times}$ plane.	
	}
\end{figure*}

In the left panel of figure \ref{htime1b} we plot $h_{+}$ (solid black curve) and $h_{\times}$ (dashed blue curve) as functions of time for a tilted torus with $B_{0}=10^{17}\,\mbox{G}$, $\Omega_{0}=100/2\pi\,\mbox{Hz}$, $\alpha=\pi/180$ and $\eta_{t}/\eta_{p}=50$.  The right hand panel shows the polarization phase portrait $[h_{+}(t),\,h_{\times}(t)]$.  The trace is not closed; it traces over an annular region, whose thickness is determined by $|\I_{2}-\I_{1}|$. 

\begin{figure}
	\begin{center}
	\includegraphics[angle=0,width=.95\columnwidth]{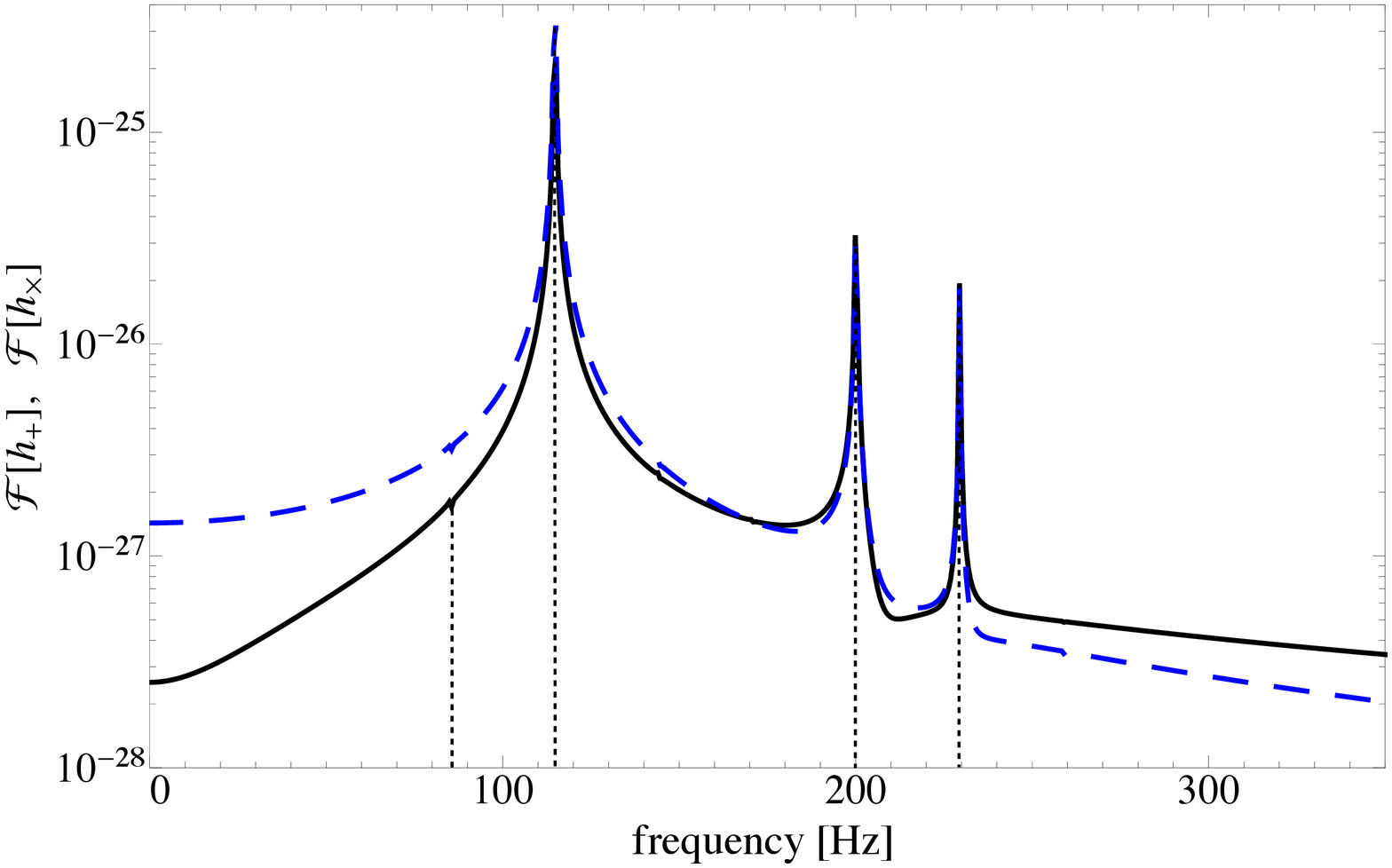}
	\end{center}
	\caption{\label{smallwobbleFourier}Fourier transform of the gravitational wave signal presented in figure \ref{htime1b}.  The solid black and dashed blue lines correspond to the plus and cross polarizations, $\mathcal{F}[h_{+}]$ and $\mathcal{F}[h_{\times}]$, respectively.  The vertical dashed lines show the position of the four spectral peaks.
	}
\end{figure}

In figure \ref{smallwobbleFourier} we plot the Fourier transform of the signal presented in figure \ref{htime1b}.  The solid black and dashed blue curves correspond to $h_{+}(f)$ and $h_{\times}(f)$ respectively.  Three unambiguous spectral lines occur at $\omega$, $2\omega/(1+\epsilon)$ and $2\omega$, as predicted analytically \cite{vandenbroeck05}.  A fourth line is also predicted at $(1-\epsilon)(1+\epsilon)^{-1}\omega$, but its amplitude is proportional to $e_{1}\I_{1}-e_{2}\I_{2}$ (cf., $e_{1}\I_{1}+e_{2}\I_{2}$ at $\omega$), i.e., it is $\sim10^{-3}$ times weaker than the line at $\omega$.  Upon closely inspecting figure \ref{smallwobbleFourier}, one can barely discern a small peak at approximately $85\,\mbox{Hz}$.

\subsection{Arbitrary wobble angle}\label{arbitrarywobble}
\begin{figure*}
	\begin{center}
	\includegraphics[angle=0,width=.95\textwidth]{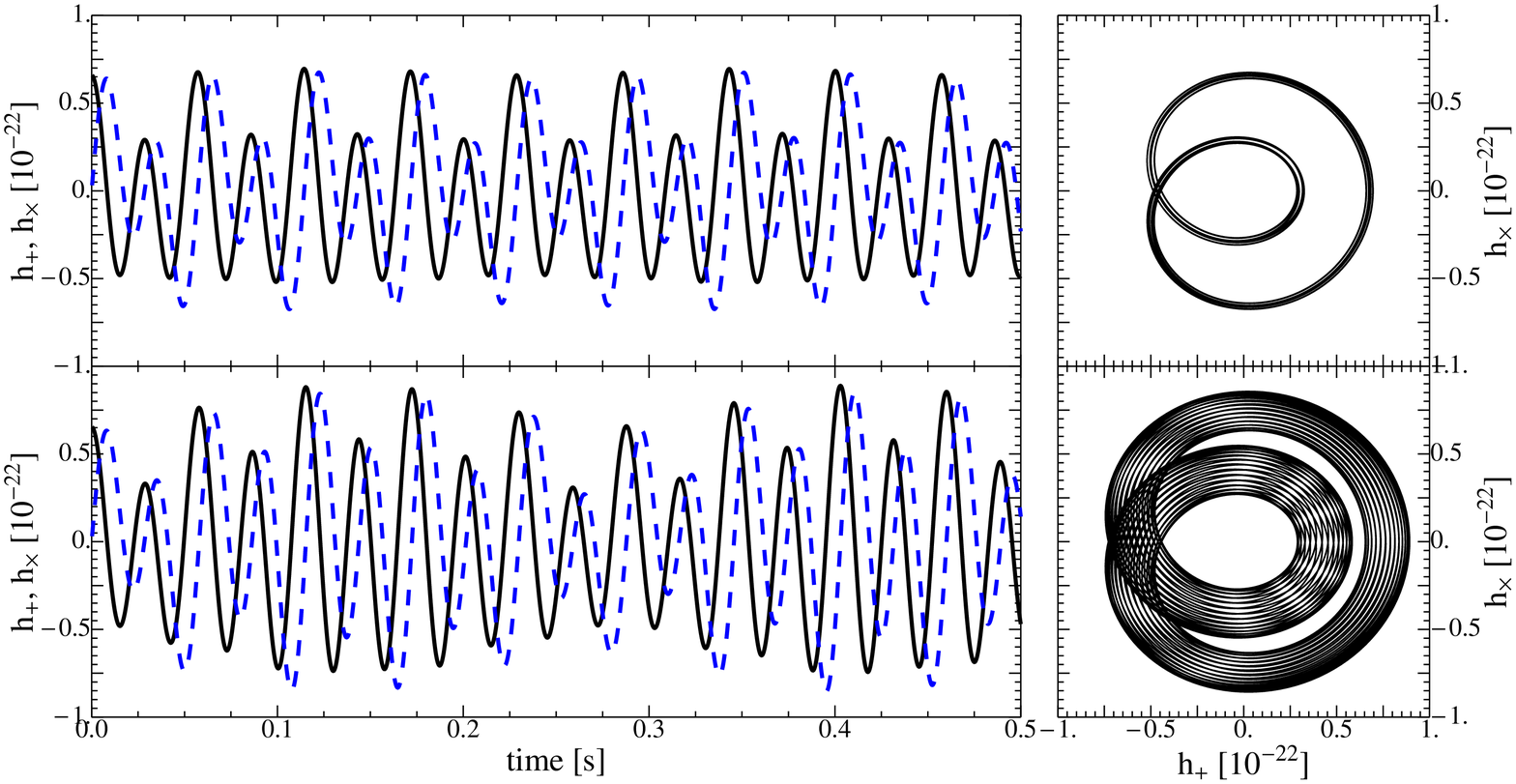}
	\end{center}
	\caption{\label{htime2}Gravitational wave signal from triaxial stars deformed by a tilted torus magnetic field.  Left panels:  Wave strain in the plus polarization $h_{+}$ (solid black curves) and cross polarization $h_{\times}$ (dashed blue curves) as functions of time for $B_{0}=10^{17}\,\mbox{G}$, $\alpha=\iota=\pi/4$ and $\eta_{t}/\eta_{p}=20$ (top panel; $\Lambda=0.93$) and $\eta_{t}/\eta_{p}=50$ (bottom panel; $\Lambda=0.67$).  Right panels:  Polarization phase portraits in the $h_{+}$-$h_{\times}$ plane.	
	}
\end{figure*}

We now consider a triaxial star with arbitrary wobble angle, $\zeta$.  As shown in figure \ref{angles}, $\zeta$ is largest when $\pi/8\lesssim\alpha\lesssim\pi/4$ and $0\leq\eta_{t}/\eta_{p}\lesssim75$.  In the two left hand panels of figure \ref{htime2} we plot $h_{+}$ (solid black curves) and $h_{\times}$ (dashed blue curve) as functions of time for stars with $\alpha=\pi/4$ and $\eta_{t}/\eta_{p}=20$ (top panel) and with $\eta_{t}/\eta_{p}=50$ (bottom panel), corresponding to $\Lambda=0.93$ and $0.67$ respectively.  The $\eta_{t}/\eta_{p}=20$ model is visually indistinguishable from the biaxial case in figure \ref{htime1}, but the $\eta_{t}/\eta_{p}=50$ model shows modulations in $h_{+,\times}(t)$ due to its larger non-axisymmetry.  In the right hand panels of figure \ref{htime2}, we plot the phase portraits for the $\eta_{t}/\eta_{p}=20$ and $50$ models.  Both differ clearly from figure \ref{htime1}; the $h_{+}$-$h_{\times}$ trajectory does not close but occupies an annulus whose thickness is determined by $|\I_{1}-\I_{2}|$.

\begin{figure*}
	\begin{center}
	\includegraphics[angle=0,width=.95\textwidth]{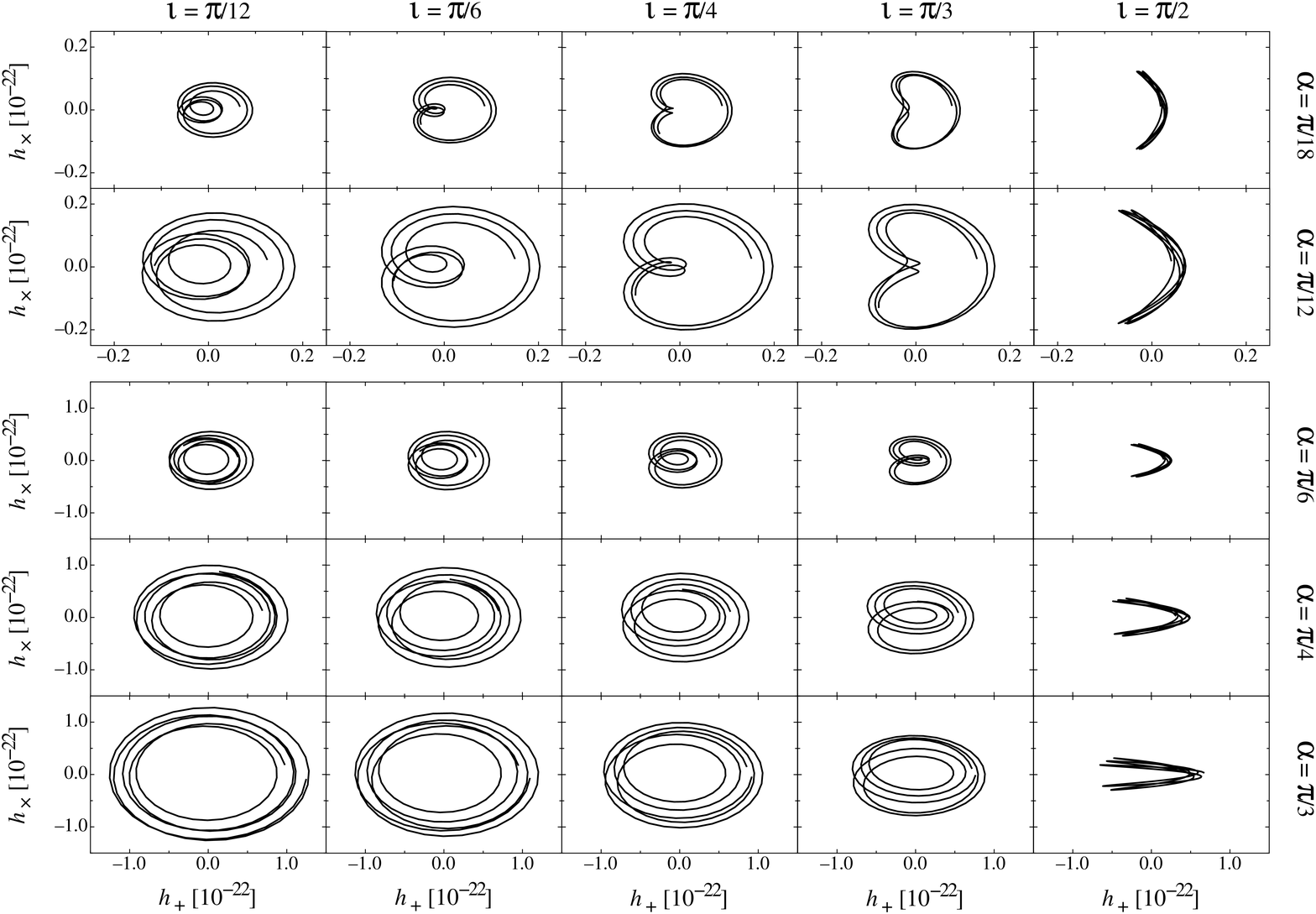}
	\end{center}
	\caption{\label{PhaseTriaxial}Polarisation phase portraits, $[h_{+}(t),\,h_{\times}(t)]$, for triaxial stars with $\eta_{t}/\eta_{p}=50$ covering a range of magnetic inclination angles ($\pi/18\le\alpha\le\pi/3$; top to bottom rows) and observer inclination angles ($\pi/12\le\iota\le\pi/3$; left to right columns).  Note the change of scale between the top two rows and the bottom three rows.  The stellar parameters are $B_{0}=10^{17}\,\mbox{G}$ and $\Omega_{0}/2\pi=100\,\mbox{Hz}$.}
\end{figure*}

The shape of the polarization phase portraits depends heavily on the magnetic and observer inclination angles, as for biaxial stars.  In figure \ref{PhaseTriaxial} we plot a grid of phase portraits for $\eta_{t}/\eta_{p}=50$ and 25 different combinations of $(\alpha,\,\iota)$.  Each panel in figure \ref{PhaseTriaxial} is evolved for the same length of time, which corresponds to a different number of cycles in phase space depending on the geometry of the field.  For example, the $(\alpha,\,\iota)=(\pi/12,\,\pi/3)$ portrait traces $\approx2.5$ cycles, whereas the $(\pi/3,\,\pi/12)$ portrait almost finishes $5.0$ complete cycles.  It is clear from figure \ref{PhaseTriaxial} that some configurations trace out distinctive phase portraits, e.g., the `bean'-like structure for $(\alpha,\,\iota)=(\pi/18,\,\pi/3)$, whereas others, like $(\pi/3,\,\pi/3)$, $(\pi/3,\,\pi/12)$ and $(\pi/12,\pi/12)$, are similar and resemble ellipses.  Hence the phase portrait is a helpful but imperfect diagnostic of the magnetic geometry.

%It is an open question as to when the trajectory closes and/or fills the annulus ergodically, as discussed in Appendix \ref{OpenClosed}\footnote{Should we include this?? It obviously needs more work, but is mainly an academic consideration).}.

\begin{figure}
	\begin{center}
	\includegraphics[angle=0,width=.95\columnwidth]{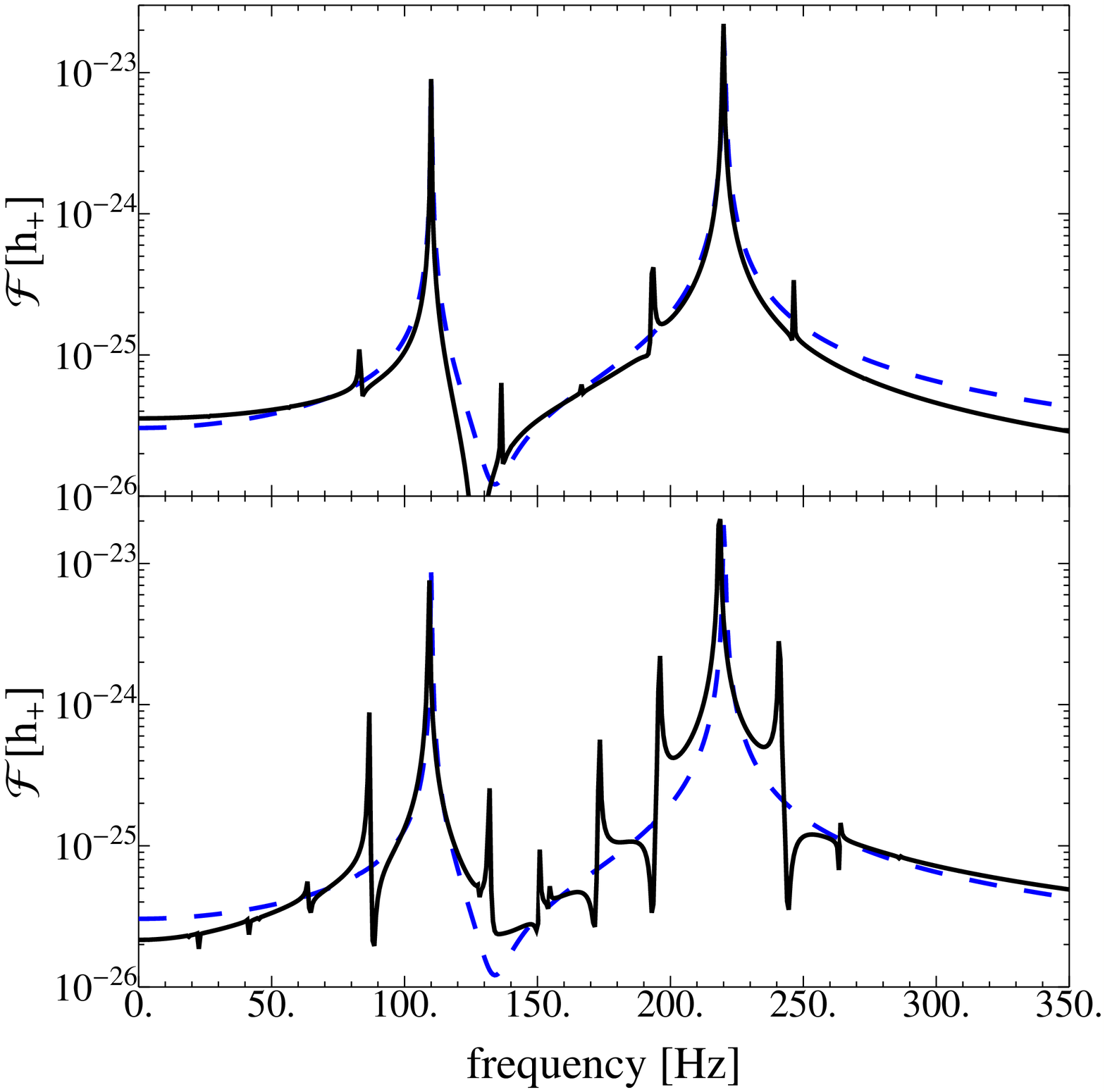}
	\end{center}
	\caption{\label{Fourier}Fourier transform of the wave strain in the plus polarization displayed in figures \ref{htime1} and \ref{htime2}.  The dashed blue curve in both panels correspond to the case with zero toroidal field (i.e. the top panel of figure \ref{htime1}) and the solid black curves correspond to the model with $\eta_{t}/\eta_{p}=20$ (top panel) and $\eta_{t}/\eta_{p}=50$ (bottom panel).  
	}
\end{figure}

In figure \ref{Fourier} we plot the Fourier transform, $\mathcal{F}[h_{+}]$, for the three models presented in figures \ref{htime1} and \ref{htime2} ($\mathcal{F}[h_{\times}]$ looks similar).  The dashed blue curve in both panels corresponds to the biaxial model from figure \ref{htime1}.  It clearly shows two peaks at $\omega$ and $2\omega$.  The solid black curves correspond to the $\eta_{t}/\eta_{p}=20$ and $50$ models in the top and bottom panels respectively.  The strongest emission occurs at angular frequencies $2\pi/T'$ and $4\pi/T'$, where $T'$ is defined in equation (\ref{2pitprime}).  
%Numerically, these evaluate to  $\omega=1.1007\Omega_{0}$, $1.0988\Omega_{0}$ and $1.0918\Omega_{0}$.  
The effect of the $\vartheta_{4}$ term in the definition of $T'$ is most evident in the bottom panel of figure \ref{Fourier}, where the peaks of the solid black and dashed blue curves at $4\pi/T'$ differ by almost $2\,\mbox{Hz}$.  Emission also occurs at angular frequencies
\begin{align}
	\frac{2\pi}{T'}+\frac{2\pi n}{T}\,\,\,\,{\rm and}\,\,\,\,
	\frac{4\pi}{T'}+\frac{2\pi n}{T},\label{frequencies}
\end{align}
for integer values of $n$, where $T$ is defined by equation (\ref{T}).  While only the spectral lines with $|n|\leq1$ are visible in the top panel of figure \ref{Fourier}, all lines with $|n|\leq3$ can be seen in the bottom panel; $n=\pm3$ requires some squinting on behalf of the reader but is confirmed under magnification.  

\begin{figure}
	\begin{center}
	\includegraphics[angle=0,width=.95\columnwidth]{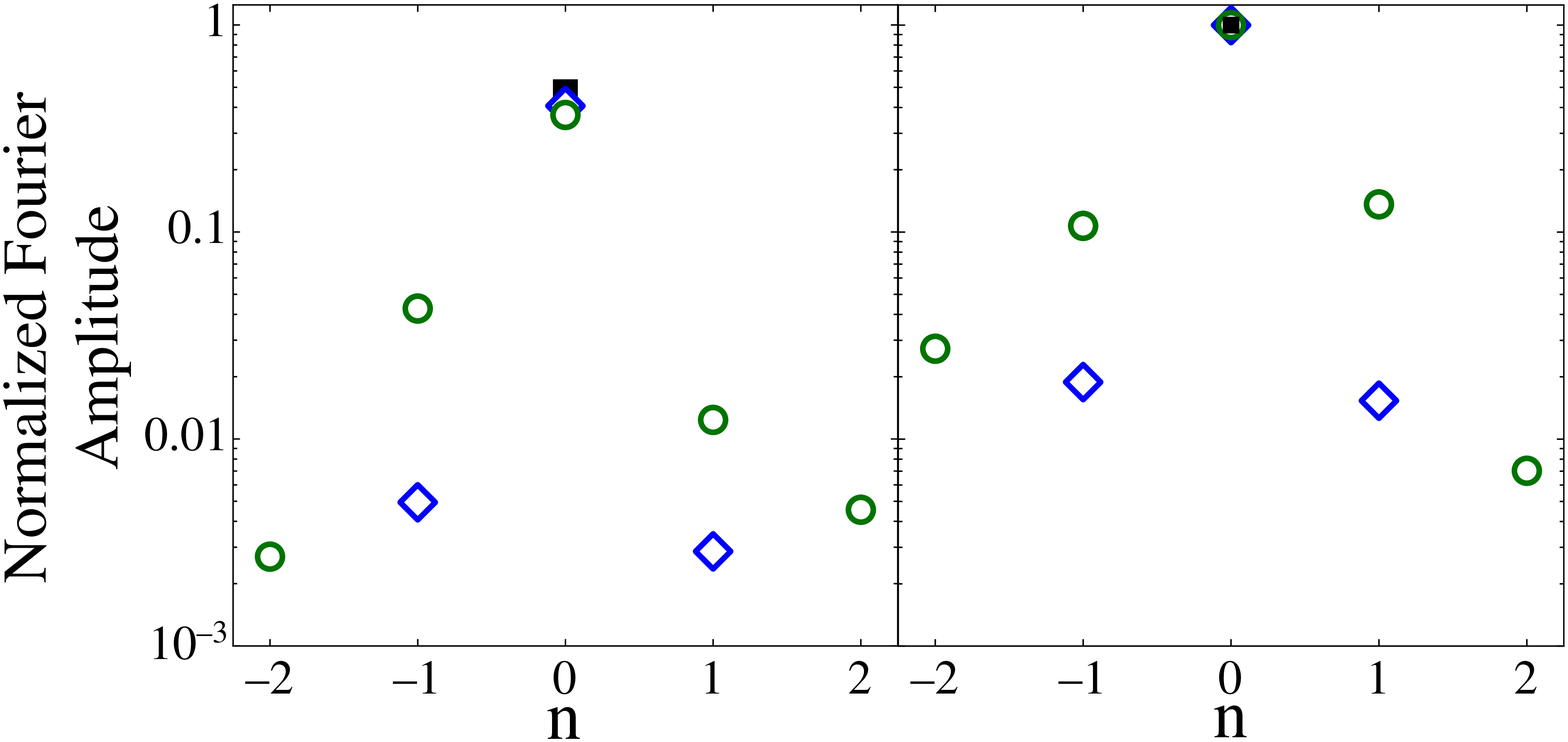}
	\end{center}
	\caption{\label{FourierRatios}Normalized Fourier peak amplitude in the plus polarization (from figure \ref{Fourier}) for $\eta_{t}=0$ (black squares), $\eta_{t}/\eta_{p}=20$ (blue diamonds) and $\eta_{t}/\eta_{p}=50$ (green circles).  Amplitudes are normalized to the largest peak (at $4\pi/T'$).  The left and right panels are the Fourier peak amplitudes around the peak at $2\pi/T'$ and $4\pi/T'$ respectively.}
\end{figure}

The relative power in the Fourier peaks can be used to determine the principal axes of inertia.  In figure \ref{FourierRatios} we plot the Fourier peak amplitudes in figure \ref{Fourier}, normalized to the tallest peak (at $4\pi/T'$), as a function of harmonic number $n$ from (\ref{frequencies}).  The left panel displays the Fourier peak amplitudes around the peak at $2\pi/T'$, and the right panel displays those around the peak at $4\pi/T'$.  The black squares, blue diamonds, and green circles correspond to $\eta_{t}=0$, $\eta_{t}/\eta_{p}=20$ and $\eta_{t}/\eta_{p}=50$ respectively.  Figure \ref{FourierRatios} shows that the amplitude of the sideband peaks (i.e., $n\neq0$) relative to the central peak ($n=0$) increases in both the left and right panels as $\eta_{t}/\eta_{p}$ increases.  Moreover, the relative amplitude of the various sidebands encodes information about $\eta_{t}/\eta_{p}$, e.g., the ratio of the $n=-1$ peaks changes as $\eta_{t}/\eta_{p}$ varies.

\begin{figure}
	\begin{center}
	\includegraphics[angle=0,width=.95\columnwidth]{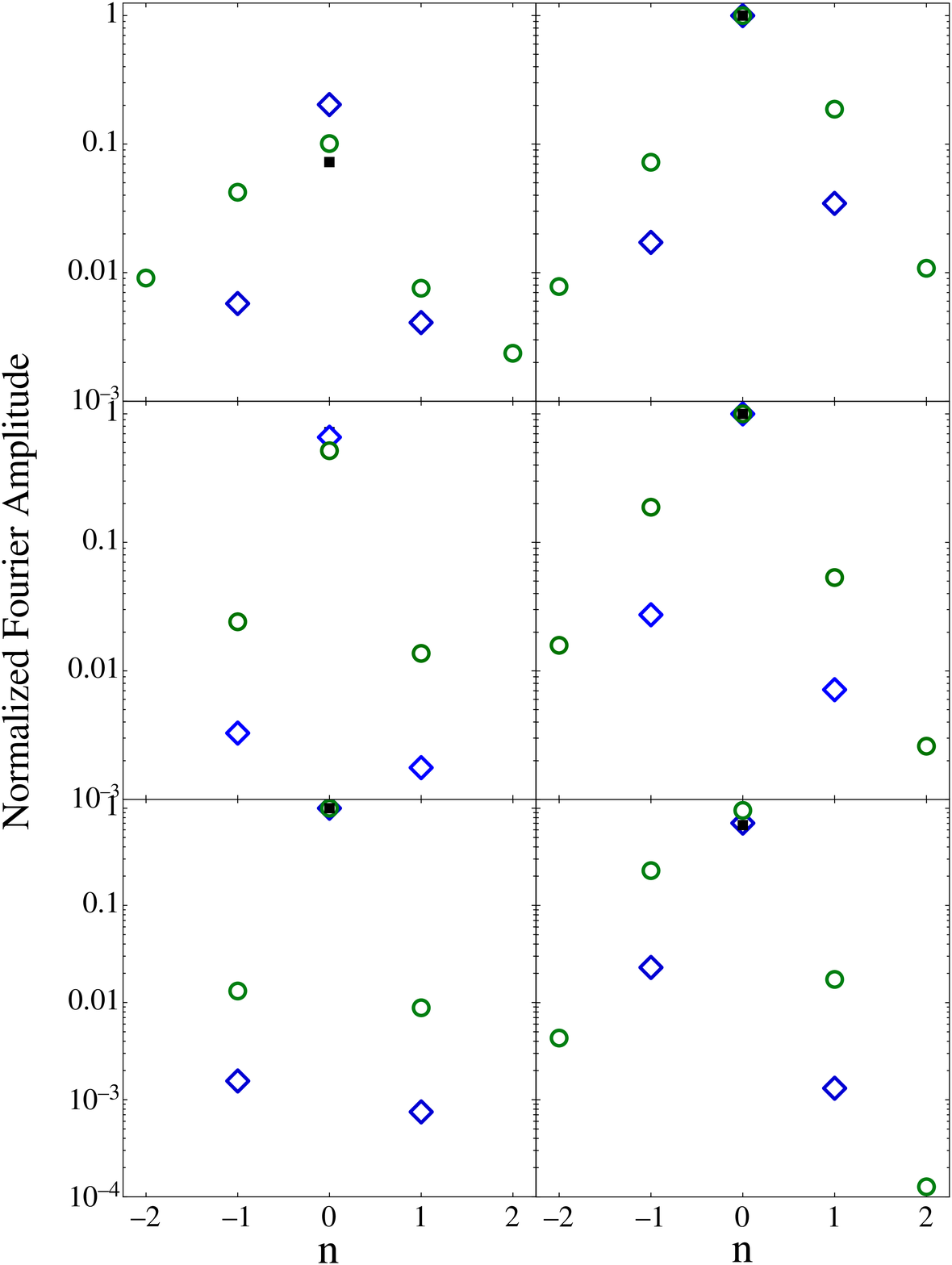}
	\end{center}
	\caption{\label{FourierRatiosall}Normalized Fourier peak amplitude in the plus polarization for stars with $\eta_{t}=0$ (black squares), $\eta_{t}/\eta_{p}=20$ (blue diamonds) and $\eta_{t}/\eta_{p}=50$ (green circles), for $\alpha=\pi/3$, $\pi/6$, and $\pi/12$ (top to bottom rows).  Amplitudes are normalized to the largest peak (at $4\pi/T'$ for $\alpha=\pi/3$ and $\pi/6$ and at $2\pi/T'$ for $\alpha=\pi/12$).  The left and right panels are the Fourier peak amplitudes around the peak at $2\pi/T'$ and $4\pi/T'$ respectively.}
\end{figure}

In figure \ref{FourierRatiosall} we show how the amplitude ratios of the Fourier peaks change as a function of the magnetic field geometry.  The panels from top to bottom correspond to $\alpha=\pi/3$, $\pi/6$ and $\pi/12$ and the Fourier peak amplitudes are normalized to the tallest peak (at $4\pi/T'$ for $\alpha=\pi/3$ and $\pi/6$ and at $2\pi/T'$ for $\alpha=\pi/12$).   The symbols are the same as for figure \ref{FourierRatios}.  Information about $\alpha$ is encoded in the ratio of power in the $2\pi/T'$ to $4\pi/T'$ peaks; the $\alpha=\pi/3$ and $\pi/4$ models are dominated by the $4\pi/T'$ peak, whereas the $2\pi/T'$ peak dominates for $\alpha=\pi/12$.  Moreover, the relative amplitude of the sidebands encodes information about both $\alpha$ and $\eta_{t}/\eta_{p}$.  

Figures \ref{FourierRatios} and \ref{FourierRatiosall} are important for inferring the internal magnetic field of a neutron star from gravitational wave observations.  We discuss how in more detail in section \ref{conclusion}.

\section{MHD Simulations of Field Winding}\label{numerical}
The tilted torus model studied in previous sections involves an idealized magnetic field suited to analytic calculations.  In this section, we motivate the model by demonstrating its similarity to magnetic fields generated by numerical simulations.  In particular, we use the three-dimensional general relativistic MHD code {\sc horizon} \cite{zink11,lasky12} to find self-consistent numerical solutions of the Einstein-Maxwell field equations with geometry resembling figure \ref{tilted}.  We deliberately limit our discussion to a single representative example.  The example serves two purposes: (i) it motivates and supports the idealized field described by equations (\ref{B})--(\ref{beta}); and (ii) it indicates qualitatively how the gravitational wave spectrum and polarization phase portrait evolve, as the internal field winds up, pointing the way to the sorts of experiments on the origin of neutron star magnetic fields that become feasible if LIGO detects the birth of a rapidly spinning protoneutron star or protomagnetar, for example.  We stress that the field configurations derived herein are highly artificial; our initial condition is a dipolar poloidal field inclined to the rotation axis, whereas a realistic field may be significantly more complicated, containing higher-order multipoles and/or a tangled component.  A full study of the gravitational wave radiation from realistically simulated magnetized stars will be the subject of future work.  

We remain agnostic as to whether the triaxiality is transient or not.  It is possible that in reality, the magnetic field ultimately moves from a non-axisymmetric state to an axisymmetric state.  Numerical simulations of Braithwaite tend to either axisymmetric \cite[e.g.,][]{braithwaite09} or non-axisymmetric configurations \cite{braithwaite08} depending on the initial conditions.  More recently, general relativistic simulations have generically evolved to non-axisymmetric configurations \cite{lasky11,lasky12,ciolfi11,ciolfi12,kiuchi11}. 

\begin{figure*}
	\begin{center}
	\includegraphics[angle=0,width=.95\textwidth]{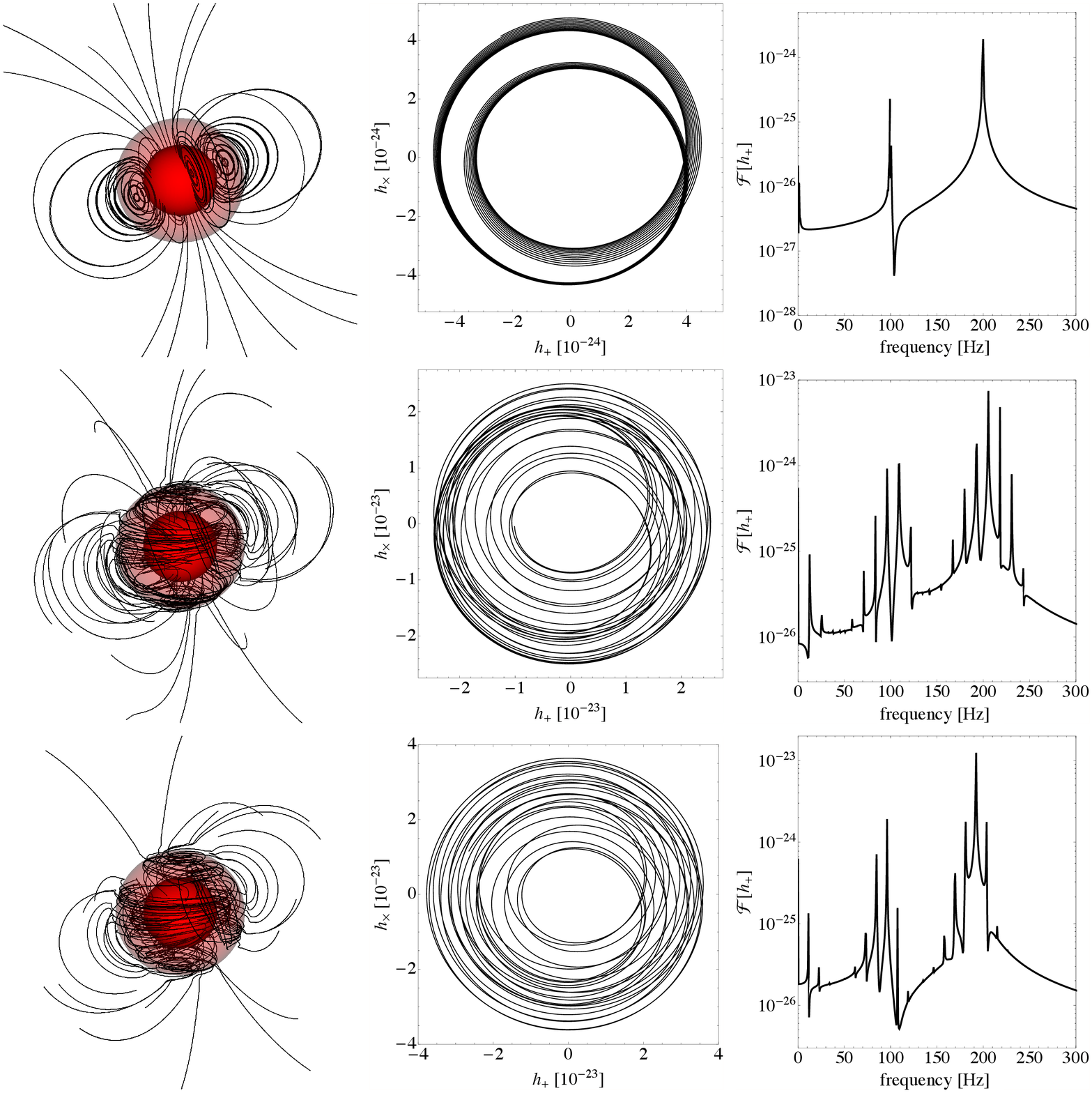}
	\end{center}
	\caption{\label{tiltedsim}Magnetic evolution of a differentially rotating star with an initially dipolar poloidal field ($B_{0}=10^{15}\,\mbox{G}$) with $\alpha=\pi/6$.  Left column: Magnetic field lines.  The opaque red and semi-transparent red iso-density surfaces correspond to $0.5\rho_{c}$ and $0.05\rho_{c}$ respectively, where $\rho_{c}$ is the central density.  The top, middle and bottom rows present snapshots after $t=0$, $1.5$ and $2.3\,\mbox{ms}$.  Central column: Polarization phase portraits.  Right column: Fourier transform of the wave strain in the plus polarization.
	}
\end{figure*}

A differentially rotating, relativistic polytrope is initialized within the publicly available Rotating Neutron Star code, {\sc rns} \cite{stergioulas95,stergioulas04}.  Differential rotation is prescribed according to the rotation law of \citet{komatsu89a,komatsu89b}, where the degree of differential rotation is set to $\hat{A}=0.8$ (see Ref. \cite{stergioulas04} for details on the rotation law) and the polar to equatorial coordinate axial ratio is $r_{p}/r_{e}=0.95$.  These parameters correspond to a star with central and equatorial angular velocities of $\Omega_{c}=1027\,\mbox{Hz}$ and $\Omega_{e}=244\,\mbox{Hz}$ respectively. We subsequently impose a dipolar poloidal magnetic field with $B_{0}=10^{15}\,\mbox{G}$ according to equation (\ref{Bp}), whose symmetry axis initially makes an angle $\alpha=\pi/6$ with the rotation axis.  The initial state is depicted in the top left panel of figure \ref{tiltedsim}, where the rotation axis points up the page and the opaque red and semi-transparent red contours are iso-density surfaces of $0.5\rho_{c}$ and $0.05\rho_{c}$ respectively, where $\rho_{c}$ is the central density.  The {\sc horizon} code then solves the general relativistic MHD evolution equations in the Cowling approximation (for details see Refs. \cite{zink11,lasky11,lasky12}).

We characterize the magnetic field using the ratio of poloidal to total field energies in the unprimed frame\footnote{It is important to note the difference between $\bar{\Lambda}$ and $\Lambda$ defined in equation (\ref{Lambda}).  The latter involves the poloidal field in the primed frame, and the toroidal field in the unprimed frame, whereas the former involves both components in the unprimed frame.  Our simulation's initial conditions are purely poloidal in the primed frame (i.e., in the frame rotated through an angle $\alpha=\pi/6$ from the rotation axis), implying $\Lambda=1$, but there is a non-zero toroidal field in the unprimed frame, which gives $\bar{\Lambda}=0.88$.  We refer to $\bar{\Lambda}$ throughout this section to conform with usage in the literature, e.g., \cite{braithwaite08,ciolfi09,mastrano11}}, $\bar{\Lambda}$.  The field initially has $\bar{\Lambda}=0.88$.  It evolves to $\bar{\Lambda}\approx0.1$ after about $0.5\,\mbox{ms}$, before oscillating between approximately $0.05$ and $0.1$ for the next few milliseconds.  The middle and bottom left panels of figure \ref{tiltedsim} display two typical snapshots at $1.5\,\mbox{ms}$ and $2.3\,\mbox{ms}$ respectively.  The figure shows that $\bar{\Lambda}$ changes as the toroidal field component winds up around the rotation axis.  The external field remains predominantly poloidal, with its axis of symmetry inclined to the rotation axis.  The middle and bottom panels resemble qualitatively the idealized magnetic field shown in figure \ref{tilted}.  Although it is difficult to visualize the poloidal component of the field inside the star, $\bar{\Lambda}\approx0.1$ implies that it remains significant energetically.

In the central and right columns of figure \ref{tiltedsim} we plot polarization phase portraits and Fourier transforms for each snapshot in the left column.  To do so, we calculate the moment-of-inertia tensor from {\sc horizon}'s output, and subtract the moment-of-inertia tensor from an otherwise identical simulation with zero magnetic field.  As the star rotates rapidly, the axisymmetric rotational bulge dominates $\I_{ij}$ yet does not contribute to $h_{+}$ and $h_{\times}$.  Subtracting an otherwise identical unmagnetized star isolates the nonaxisymmetric part of $\I_{ij}$ and hence $h_{+}$ and $h_{\times}$.  We multiply $\I_{ij}$ by $10^{4}$ and rescale it onto the model evaluated in previous sections (i.e., a uniformly rotating star with $\Omega_{0}/2\pi=100\,\mbox{Hz}$, $\iota=\pi/4$ and $d=1\,\mbox{kpc}$)\footnote{To bring this section into line with preceding sections, we artificially strengthen the magnetic field by two orders of magnitude, corresponding to four orders of magnitude in the moment of inertia tensor.}.

The $t=0$ snapshot (top row of figure \ref{tiltedsim}) contains no magnetic deformation.  We therefore evolve the star $0.025\,\mbox{ms}$ before calculating the polarization phase portrait and Fourier transform.  At this point the star is almost biaxial; the Fourier transform is visually indistinguishable from that from a biaxial star, however the polarization phase portrait is distinctively triaxial, i.e., the trajectory does not close.  At $t=1.5\,\mbox{ms}$ and $2.3\,\mbox{ms}$ a strong toroidal component is wound up around the rotation axis and the phase portraits show that the star is triaxial.  The Fourier transforms show complicated spectra, with harmonics of orders $|n|\le4$ visible in the $t=1.5\,\mbox{ms}$ snapshot.  Moreover, the relative amplitude of the various peaks is seen to evolve between $t=1.5\,\mbox{ms}$ and $2.3\,\mbox{ms}$.  For example, while the relative size of the $n=0$ peaks remains relatively constant, the two $n=1$ peaks are significantly weaker after $2.3\,\mbox{ms}$.  This hints that the wobble angle remains constant between the two snapshots, while the toroidal-poloidal ratio changes.  When an instrument like Advanced LIGO observes gravitational waves from the birth of a neutron star, the evolution of the relative Fourier peak amplitudes seen in figure \ref{tiltedsim} may allow one to reconstruct the evolution of the magnetic field.

The snapshots in figure \ref{tiltedsim} motivate qualitatively the use of the analytic model in section \ref{model}.  However, there is much still to be explored regarding these new field configurations.  One significant difference between the snapshots and figure \ref{tilted} is the distribution of toroidal field.  The numerical simulations wind up a toroidal component throughout the star, whereas the analytic model confines the toroidal component to a region near the equator in the unprimed frame.  This toroidal field is approximately constant throughout the volume of the star; note that the analytic model has the toroidal field occupying approximately $20\%$ of the volume.  Moreover, the stability of these fields is still an open question, given we evolve the system for only $\sim10$ Alfv\'en crossing times.  These and related issues, e.g., what effect the initial magnetic field distribution has on the steady state, will be explored in detail in subsequent work.

\section{Conclusion}\label{conclusion}
In this article, we construct `tilted torus' magnetic field configurations, whose toroidal and poloidal axes of symmetry are misaligned.  The toroidal component is assumed to wind up around the rotation axis following the action of $r$-modes or differential rotation in the protoneutron star.  The poloidal component, whose symmetry axis makes an angle $\alpha$ with the rotation axis, is a fossil of the protoneutron star's field.  A tilted torus deforms the star triaxially, unlike a twisted torus, which produces a biaxial deformation.  We take the first steps towards analysing the gravitational wave signal to see what can, and cannot, be inferred about the magnetic field geometry from future observations.

To aid in the above task, we develop a new diagnostic: polarization phase portraits in the $h_{+}$-$h_{\times}$ plane.  Biaxial stars trace closed loops in the $h_{+}$-$h_{\times}$ plane; see the right hand panel of figures \ref{htime1} and \ref{PhaseBiaxial}.  Triaxial stars trace an open path which wanders within an annulus whose thickness depends on $|\I_{1}-\I_{2}|$; see right-hand panels of figures \ref{htime2} and \ref{PhaseTriaxial}.  Hence the phase portrait is a promising way to discriminate between twisted torus and tilted torus magnetic configurations.  Figures \ref{PhaseBiaxial} and \ref{PhaseTriaxial} show that the magnetic and observer inclination angles, $\alpha$ and $\iota$ respectively, can be inferred from the polarization phase portrait in certain cases (e.g., for a biaxial star, or a triaxial star with a `bean-shaped' portrait) but not in others, e.g., the portraits for $(\alpha,\,\iota)=(\pi/18,\,\pi/12)$, $(\pi/3,\,\pi/12)$ and $(\pi/3,\,\pi/3)$ closely resemble one another.

%In reality though, if the deviation from biaxiality is small enough that the phase portraits vary a small amount per stellar spin period, then the phase portraits will show characteristic traits akin to that seen in the biaxial phase portraits.  For example, the portrait in the top panel of figure \ref{tiltedsim} will be more useful for determining the stars principal moments than the middle and bottom panels.

The gravitational wave spectrum from the tilted torus exhibits emission at the angular frequencies given by (\ref{frequencies}).  The Fourier peaks with $n=0$ have the greatest amplitude.  For a biaxial star, they lie at $\omega$ and $2\omega$; for a triaxial star, they are shifted by the amount in the final term in equation (\ref{2pitprime}), as much as $1\%$ for the models in figure \ref{Fourier}.  The frequencies of the $n=0$ peaks do not provide information about the magnetic geometry, but their relative amplitude does (see below).  For triaxial stars, the $n=0$ peaks are straddled by weaker spectral lines displaced by $2\pi n/T$, where $T$ is defined in equation (\ref{T}).

One consequence of the results presented here is the ability to discern twisted and tilted torus magnetic field geometries.  A star deformed by a twisted torus field emits at two frequencies, and the polarization phase portrait traces a closed curve.  Stars deformed by tilted tori emit more than two frequencies and their phase portraits do not trace closed curves (see section \ref{triaxial}).  Gravitational wave observations can therefore discern twisted and tilted torus magnetic configurations in principle, even where low signal-to-noise prohibits identification of the $n\neq0$ peaks in the Fourier spectrum.

Given gravitational wave observations of a triaxial star, how can we infer its magnetic field?  Figures \ref{FourierRatios} and \ref{FourierRatiosall} show that geometric information is encoded in the line-amplitude ratios.  For example,
\begin{itemize}
	\item As $\alpha$ decreases, the ratio of power in the $2\pi/T'$ to $4\pi/T'$ (i.e., $n=0$) peaks increases; the $4\pi/T'$ peak dominates for $\alpha\lesssim\pi/12$, whereas the $2\pi/T'$ peak dominates for $\alpha\gtrsim\pi/12$.
	\item The amplitude of the sidebands ($n\neq0$) relative to the central peak ($n=0$) increases as the toroidal-poloidal energy ratio $\eta_{t}/\eta_{p}$ increases. 
	\item The relative amplitudes of the sidebands encodes information about $\alpha$ and $\eta_{t}/\eta_{p}$, e.g., the ratio of the two $n=-1$ peaks changes systematically as both $\alpha$ and $\eta_{t}/\eta_{p}$ vary.
\end{itemize}
Inferring the internal magnetization of a neutron star from gravitational wave observations therefore requires careful comparisons of the observed spectral lines with a collection of templates like those in figures \ref{FourierRatios} and \ref{FourierRatiosall}.  We note that realistic magnetic fields may well be more complicated than those presented here, but $\I_{ij}$ has only three eigenvalues for all field structures, so the features in figures \ref{FourierRatios} and \ref{FourierRatiosall} do not change qualitatively, even though their interpretation does.

%Figures \ref{FourierRatios} and \ref{FourierRatiosall} show that the magnetic field geometry is encoded in the relative power of the different spectral peaks.  For example, as $\alpha$ decreases, the ratio of power in the main $2\pi/T'$ to $4\pi/T'$ (i.e., $n=0$) peaks increase; the $4\pi/T'$ peak dominates for $\alpha\lesssim\pi/12$, but becomes weaker than the $2\pi/T'$ spectral peak for $\alpha=\pi/12$.  The relative strength of the toroidal and poloidal components has a more significant effect on the relative power in the sideband spectral peaks.  For example, purely poloidal models ($\eta_{t}=0$, where $\eta_{t}$ defines the relative strength of the toroidal component with respect to the poloidal component) only have two emission frequencies, whereas the amplitude of the sideband peaks ($n\neq0$) relative to the central peaks ($n=0$) increase as $\eta_{t}/\eta_{p}$ increases.  The relative amplitude of the sideband peaks with one another also encodes information about the magnetic field topology.  For example, the ratio of the two $n=-1$ peaks changes as both $\alpha$ and $\eta_{t}/\eta_{p}$ vary.  Inferring a stars magnetic field geometry from gravitational wave observations therefore requires careful comparisons of the observed spectral peaks with more detailed versions of figures \ref{FourierRatios} and \ref{FourierRatiosall}.  In particular, a collection of templates with the relative amplitude and frequencies of spectral lines for different magnetic field geometries and stellar spin orientations will be required.

It is still an open question whether the configuration in section \ref{model} is generically stable.  \citet{akgun13} recently showed that the $\alpha=0$ case for a similar field configuration\footnote{In Ref. \cite{akgun13}, the toroidal component was defined as $\beta(\gamma)=(\gamma-1)^{2}$ for $\gamma\ge1$, compared to $\beta(\gamma)=\gamma-1$ in our model; see equation (\ref{beta}) and section \ref{model}.} is stable for reasonable values of $\eta_{t}/\eta_{p}$, qualitatively supporting previous numerical work \cite{braithwaite09}.  However, a new type of instability may occur that restores the field to an axisymmetric configuration or rearranges it completely, e.g., \cite{vigelius08}.  For newborn magnetars it is interesting to ask whether such instabilities saturate before or after the toroidal component winds up.  Numerical simulations of magnetic fields in slowly rotating stars evolve to axisymmetric \cite[e.g.,][]{braithwaite06,braithwaite09} and non-axisymmetric \cite{braithwaite08,lasky11,ciolfi11,kiuchi11} configurations depending on a number of different factors including the initial conditions and the degree of stratification.  It is therefore unclear whether one expects newly born neutron stars to be biaxial or triaxial, nor whether the triaxiality is transient or persistent.  The gravitational wave diagnostics developed in this paper help prepare to answer this question.

The detectability of gravitational waves from a neutron star with a tilted torus depends sensitively on how long the magnetic quadrupole lasts.  If the non-axisymmetric field is transitory, surviving only until an instability acts to symmetrize the field, then the detectability is significantly diminished; the signal-to-noise ratio (SNR) scales as $T^{1/2}$, where $T$ is the lesser of the observation time and the emitting time \cite[e.g.,][]{sathyaprakash09}.  If a dynamical instability symmetrizes the field on the Alfv\'en timescale, then $T$ is of order tens to hundreds of seconds.  If the non-axisymmetric field configuration is stable, $T$ may be months to years.  

The SNR for a protomagnetar observed with Advanced LIGO in the frequency range $500$--$2000\,{\rm Hz}$ is \cite{dallosso09}
\begin{align}
	{\rm SNR}=&3\left(\frac{\left<B_{t}\right>}{2\times10^{16}\,{\rm G}}\right)^{2}\left(\frac{B_{p}}{10^{14}\,{\rm G}}\right)^{-1}\left(\frac{d}{20\,{\rm Mpc}}\right)^{-1}\notag\\
	& \times\left[\ln\left(\frac{a^{2}+f_{f}^{2}}{a^{2}+f_{i}^{2}}\right)+2\ln\left(\frac{f_{i}}{f_{f}}\right)\right]^{1/2},\label{SNR}
\end{align}
where $a^{2}=2K_{\rm em}/\pi^{2}K_{\rm gw}$, $K_{\rm em}=B_{p}^{2}R^{6}/(3Ic^{3})$, $K_{\rm gw}=128GI\epsilon^{2}/(5c^{5})$, $\left<B_{t}\right>$ is the average internal field strength and $f_{i}$, $f_{f}$ are the initial and final spin frequencies at $t=0$ and $T$ respectively.  The term in the final square brackets of equation (\ref{SNR}) expresses the $T^{1/2}$ scaling in terms of the spin down of the neutron star due to both gravitational wave and electromagnetic torques (see Ref. \cite{dallosso09} for details).  As a representative example, consider a neutron star in the Virgo cluster (i.e., $d\approx20\,{\rm Mpc}$) born with $B_{p}=10^{14}\,{\rm G}$, $\left<B_{t}\right>=2\times10^{16}\,{\rm G}$ and initial spin period of $10\,{\rm ms}$.  If the triaxiality survives for $\sim1\,{\rm s}$, one finds ${\rm SNR}\sim10^{-2}$.  On the other hand, a persistent magnetic field that allows one month of observations provides a border-line case for detection with ${\rm SNR}\approx3$.

%Reconstructing a stars magnetic field geometry from gravitational wave observations requires careful comparisons of the observed spectral peaks with more detailed versions of figures \ref{FourierRatios} and \ref{FourierRatiosall}.  A collection of templates with the relative amplitude and frequencies of spectral lines for different magnetic field and spin geometries will be required.  

The magnetic field geometry in section \ref{model} is one convenient analytic generalization of the twisted torus fields popularized by recent state-of-the-art numerical simulations \cite{geppert06,braithwaite06a,braithwaite06b,braithwaite06,braithwaite07,braithwaite08,braithwaite09,kiuchi11,lasky11,ciolfi12,lasky12}.  A more thorough analysis of other possible non-axisymmetric magnetic field configurations is required.  Analytic investigations should include more realistic density profiles, relativistic gravity and gravitational perturbations (i.e., not the Cowling approximation) as discussed in sections \ref{model} and \ref{perturb}.  For more realistic models, this requires numerical simulations that include angular velocity profiles from the end-state of three-dimensional core-collapse simulations (e.g., \cite{ott06}) or the merger of two neutron stars \cite{giacomazzo13}, including three-dimensional neutrino transport to power turbulent convection.  Understanding possible magnetic field configurations on longer timescales includes studying higher-order multipoles \cite{mastrano13}, superfluidity and superconductivity \cite{glampedakis12,glampedakis12a,lander12b,lander12c,mastrano12} and the role of the crust (e.g., \cite{pons09,gabler11,gabler12,vigano12,vigano13,gabler13b}).

%The magnetic fields studied here are idealized.  We only consider a dipolar poloidal component; higher multipoles are the subject of an accompanying article \cite{mastrano13}.  Moreover, we isolate the toroidal component to a doughnut near the equator, whereas preliminary numerical simulations of differentially rotating fluid balls (see section \ref{numerical})
%using the GRMHD {\sc horizon} code \cite{zink11,lasky12} 
%indicate that it penetrates the entire differentially rotating region.  A thorough analysis starts with angular velocity profiles from the end-state of three-dimensional core-collapse simulations (e.g., \cite{ott06}), including three-dimensional neutrino transport to power turbulent convection and hence a dynamo.

%Although the paper is primarily motivated by newly born stars, it may also be relevant for young radio pulsars like the Crab and Vela.  If the cores of such objects are in colour superconducting states, their ellipticity due to magnetic deformations may be enhanced, to the point where they are borderline detectable by Advanced LIGO \cite{glampedakis12}.  The study of magnetic field configurations in superconducting stars is still in its infancy \cite{glampedakis12a,lander12b,lander12c,mastrano12}. 
%--- the Lorentz force takes a different form 

Precession has been verified in only one radio pulsar \cite{stairs00}, although numerous other results hint at free precession, including recent observations of a helical structure in the jet emanating from Vela \cite{durant13}.  Observations of precession could provide important clues into the internal state of neutron stars beyond their magnetic field.  For example, the absence of precession may hint at a superfluid interior (e.g., \cite{jones01}), although coupling between crust and core significantly complicates interpretations of any results (e.g., \cite{levin04}).  Finally, if the core superrotates with respect to the crust \cite{melatos12}, the core may precess even while the crust does not.  Core superrotation may also drive ongoing magnetic activity, so that the gravitational wave signature from the magnetic deformation is more complicated than the calculations in this paper imply.

\acknowledgments
We are grateful to Alpha Mastrano and Sam Lander for valuable discussions, and to Anton Tarasenko for an early implementation of the moment of inertia calculations within the {\sc horizon} code.  PL is especially grateful to Burkhard Zink and Kostas Kokkotas for work and discussions related to the {\sc horizon} code.  We thank the anonymous referee for insightful feedback that has improved the manuscript.  This work is supported by an Australian Research Council Discovery Project grant (DP110103347).  PL is supported by an internal University of Melbourne Early Career Researcher Grant.  The numerical simulation was performed on the Multi-modal Australian ScienceS Imaging and Visualisation Environment (MASSIVE; www.massive.org.au) through an award under the Merit Allocation Scheme on the NCI National Facility at the ANU.

\bibliography{Oblique_Rotators.V5}

\begin{thebibliography}{72}
\expandafter\ifx\csname natexlab\endcsname\relax\def\natexlab#1{#1}\fi
\expandafter\ifx\csname bibnamefont\endcsname\relax
  \def\bibnamefont#1{#1}\fi
\expandafter\ifx\csname bibfnamefont\endcsname\relax
  \def\bibfnamefont#1{#1}\fi
\expandafter\ifx\csname citenamefont\endcsname\relax
  \def\citenamefont#1{#1}\fi
\expandafter\ifx\csname url\endcsname\relax
  \def\url#1{\texttt{#1}}\fi
\expandafter\ifx\csname urlprefix\endcsname\relax\def\urlprefix{URL }\fi
\providecommand{\bibinfo}[2]{#2}
\providecommand{\eprint}[2][]{\url{#2}}

\bibitem[{\citenamefont{{{Abbott}, B.~P. and {Abbott}, R. and {Adhikari}, R.
  and {Ajith}, P. and {Allen}, B. and {Allen}, G. and {Amin}, R.~S. and
  {Anderson}, S.~B. and {Anderson}, W.~G. and {Arain}, M.~A. and et
  al.}}(2009)}]{abbott09a}
\bibinfo{author}{\bibnamefont{{{Abbott}, B.~P. and {Abbott}, R. and {Adhikari},
  R. and {Ajith}, P. and {Allen}, B. and {Allen}, G. and {Amin}, R.~S. and
  {Anderson}, S.~B. and {Anderson}, W.~G. and {Arain}, M.~A. and et al.}}},
  \bibinfo{journal}{Rep. Prog. Phys.} \textbf{\bibinfo{volume}{72}},
  \bibinfo{pages}{076901} (\bibinfo{year}{2009}).

\bibitem[{\citenamefont{Duncan and Thompson}(1992)}]{duncan92}
\bibinfo{author}{\bibfnamefont{R.~C.} \bibnamefont{Duncan}} \bibnamefont{and}
  \bibinfo{author}{\bibfnamefont{C.}~\bibnamefont{Thompson}},
  \bibinfo{journal}{Astrophys. J.} \textbf{\bibinfo{volume}{392}},
  \bibinfo{pages}{L9} (\bibinfo{year}{1992}).

\bibitem[{\citenamefont{Thompson and Duncan}(1993)}]{thompson93}
\bibinfo{author}{\bibfnamefont{C.}~\bibnamefont{Thompson}} \bibnamefont{and}
  \bibinfo{author}{\bibfnamefont{R.~C.} \bibnamefont{Duncan}},
  \bibinfo{journal}{Astrophys. J.} \textbf{\bibinfo{volume}{408}},
  \bibinfo{pages}{194} (\bibinfo{year}{1993}).

\bibitem[{\citenamefont{Ioka}(2001)}]{ioka01}
\bibinfo{author}{\bibfnamefont{K.}~\bibnamefont{Ioka}}, \bibinfo{journal}{Mon.
  Not. R. Astron. Soc.} \textbf{\bibinfo{volume}{327}}, \bibinfo{pages}{639}
  (\bibinfo{year}{2001}).

\bibitem[{\citenamefont{Palomba}(2001)}]{palomba01}
\bibinfo{author}{\bibfnamefont{C.}~\bibnamefont{Palomba}},
  \bibinfo{journal}{A\&A} \textbf{\bibinfo{volume}{367}}, \bibinfo{pages}{525}
  (\bibinfo{year}{2001}).

\bibitem[{\citenamefont{Stella et~al.}(2005)\citenamefont{Stella, Dall'Osso,
  Israel, and Vecchio}}]{stella05}
\bibinfo{author}{\bibfnamefont{L.}~\bibnamefont{Stella}},
  \bibinfo{author}{\bibfnamefont{S.}~\bibnamefont{Dall'Osso}},
  \bibinfo{author}{\bibfnamefont{G.~L.} \bibnamefont{Israel}},
  \bibnamefont{and} \bibinfo{author}{\bibfnamefont{A.}~\bibnamefont{Vecchio}},
  \bibinfo{journal}{Astrophys. J.} \textbf{\bibinfo{volume}{634}},
  \bibinfo{pages}{L165} (\bibinfo{year}{2005}).

\bibitem[{\citenamefont{Dall'Osso et~al.}(2009)\citenamefont{Dall'Osso, Shore,
  and Stella}}]{dallosso09}
\bibinfo{author}{\bibfnamefont{S.}~\bibnamefont{Dall'Osso}},
  \bibinfo{author}{\bibfnamefont{S.~N.} \bibnamefont{Shore}}, \bibnamefont{and}
  \bibinfo{author}{\bibfnamefont{L.}~\bibnamefont{Stella}},
  \bibinfo{journal}{Mon. Not. R. Astron. Soc.} \textbf{\bibinfo{volume}{398}},
  \bibinfo{pages}{1869} (\bibinfo{year}{2009}).

\bibitem[{\citenamefont{Cutler}(2002)}]{cutler02}
\bibinfo{author}{\bibfnamefont{C.}~\bibnamefont{Cutler}},
  \bibinfo{journal}{Phys. Rev. D} \textbf{\bibinfo{volume}{66}},
  \bibinfo{pages}{084025} (\bibinfo{year}{2002}).

\bibitem[{\citenamefont{Glampedakis
  et~al.}(2012{\natexlab{a}})\citenamefont{Glampedakis, Jones, and
  Samuelsson}}]{glampedakis12}
\bibinfo{author}{\bibfnamefont{K.}~\bibnamefont{Glampedakis}},
  \bibinfo{author}{\bibfnamefont{D.~I.} \bibnamefont{Jones}}, \bibnamefont{and}
  \bibinfo{author}{\bibfnamefont{L.}~\bibnamefont{Samuelsson}},
  \bibinfo{journal}{Phys. Rev. Lett.} \textbf{\bibinfo{volume}{109}},
  \bibinfo{pages}{081103} (\bibinfo{year}{2012}{\natexlab{a}}).

\bibitem[{\citenamefont{{Abbott} et~al.}(2008)\citenamefont{{Abbott}, {Abbott},
  {Adhikari}, {Ajith}, {Allen}, {Allen}, {Amin}, {Anderson}, {Anderson},
  {Arain} et~al.}}]{abbott08c}
\bibinfo{author}{\bibfnamefont{B.}~\bibnamefont{{Abbott}}},
  \bibinfo{author}{\bibfnamefont{R.}~\bibnamefont{{Abbott}}},
  \bibinfo{author}{\bibfnamefont{R.}~\bibnamefont{{Adhikari}}},
  \bibinfo{author}{\bibfnamefont{P.}~\bibnamefont{{Ajith}}},
  \bibinfo{author}{\bibfnamefont{B.}~\bibnamefont{{Allen}}},
  \bibinfo{author}{\bibfnamefont{G.}~\bibnamefont{{Allen}}},
  \bibinfo{author}{\bibfnamefont{R.}~\bibnamefont{{Amin}}},
  \bibinfo{author}{\bibfnamefont{S.~B.} \bibnamefont{{Anderson}}},
  \bibinfo{author}{\bibfnamefont{W.~G.} \bibnamefont{{Anderson}}},
  \bibinfo{author}{\bibfnamefont{M.~A.} \bibnamefont{{Arain}}},
  \bibnamefont{et~al.}, \bibinfo{journal}{Astrophys. J.}
  \textbf{\bibinfo{volume}{683}}, \bibinfo{pages}{L45} (\bibinfo{year}{2008}).

\bibitem[{\citenamefont{Bonazzola and Gourgoulhon}(1996)}]{bonazzola96}
\bibinfo{author}{\bibfnamefont{S.}~\bibnamefont{Bonazzola}} \bibnamefont{and}
  \bibinfo{author}{\bibfnamefont{E.}~\bibnamefont{Gourgoulhon}},
  \bibinfo{journal}{A\&A} \textbf{\bibinfo{volume}{312}}, \bibinfo{pages}{675}
  (\bibinfo{year}{1996}).

\bibitem[{\citenamefont{Kruskal and Schwarzschild}(1954)}]{kruskal54}
\bibinfo{author}{\bibfnamefont{M.}~\bibnamefont{Kruskal}} \bibnamefont{and}
  \bibinfo{author}{\bibfnamefont{M.}~\bibnamefont{Schwarzschild}},
  \bibinfo{journal}{Proc. Roy. Soc. A} \textbf{\bibinfo{volume}{223}},
  \bibinfo{pages}{348} (\bibinfo{year}{1954}).

\bibitem[{\citenamefont{Tayler}(1957)}]{tayler57}
\bibinfo{author}{\bibfnamefont{R.~J.} \bibnamefont{Tayler}},
  \bibinfo{journal}{Proc. Phys. Soc. B} \textbf{\bibinfo{volume}{70}},
  \bibinfo{pages}{31} (\bibinfo{year}{1957}).

\bibitem[{\citenamefont{Tayler}(1973)}]{tayler73}
\bibinfo{author}{\bibfnamefont{R.~J.} \bibnamefont{Tayler}},
  \bibinfo{journal}{Mon. Not. R. Astron. Soc.} \textbf{\bibinfo{volume}{161}},
  \bibinfo{pages}{365} (\bibinfo{year}{1973}).

\bibitem[{\citenamefont{Wright}(1973)}]{wright73}
\bibinfo{author}{\bibfnamefont{G.~A.~E.} \bibnamefont{Wright}},
  \bibinfo{journal}{Mon. Not. R. Astron. Soc.} \textbf{\bibinfo{volume}{162}},
  \bibinfo{pages}{339} (\bibinfo{year}{1973}).

\bibitem[{\citenamefont{Markey and Tayler}(1973)}]{markey73}
\bibinfo{author}{\bibfnamefont{P.}~\bibnamefont{Markey}} \bibnamefont{and}
  \bibinfo{author}{\bibfnamefont{R.~J.} \bibnamefont{Tayler}},
  \bibinfo{journal}{Mon. Not. R. Astron. Soc.} \textbf{\bibinfo{volume}{163}},
  \bibinfo{pages}{77} (\bibinfo{year}{1973}).

\bibitem[{\citenamefont{Markey and Tayler}(1974)}]{markey74}
\bibinfo{author}{\bibfnamefont{P.}~\bibnamefont{Markey}} \bibnamefont{and}
  \bibinfo{author}{\bibfnamefont{R.~J.} \bibnamefont{Tayler}},
  \bibinfo{journal}{Mon. Not. R. Astron. Soc.} \textbf{\bibinfo{volume}{168}},
  \bibinfo{pages}{505} (\bibinfo{year}{1974}).

\bibitem[{\citenamefont{Geppert and Rheinhardt}(2006)}]{geppert06}
\bibinfo{author}{\bibfnamefont{U.}~\bibnamefont{Geppert}} \bibnamefont{and}
  \bibinfo{author}{\bibfnamefont{M.}~\bibnamefont{Rheinhardt}},
  \bibinfo{journal}{A\&A} \textbf{\bibinfo{volume}{456}}, \bibinfo{pages}{639}
  (\bibinfo{year}{2006}).

\bibitem[{\citenamefont{Braithwaite}(2006)}]{braithwaite06a}
\bibinfo{author}{\bibfnamefont{J.}~\bibnamefont{Braithwaite}},
  \bibinfo{journal}{A\&A} \textbf{\bibinfo{volume}{453}}, \bibinfo{pages}{687}
  (\bibinfo{year}{2006}).

\bibitem[{\citenamefont{Braithwaite and Nordlund}(2006)}]{braithwaite06b}
\bibinfo{author}{\bibfnamefont{J.}~\bibnamefont{Braithwaite}} \bibnamefont{and}
  \bibinfo{author}{\bibfnamefont{A.}~\bibnamefont{Nordlund}},
  \bibinfo{journal}{A\&A} \textbf{\bibinfo{volume}{450}}, \bibinfo{pages}{1077}
  (\bibinfo{year}{2006}).

\bibitem[{\citenamefont{Braithwaite and Spruit}(2006)}]{braithwaite06}
\bibinfo{author}{\bibfnamefont{J.}~\bibnamefont{Braithwaite}} \bibnamefont{and}
  \bibinfo{author}{\bibfnamefont{H.~C.} \bibnamefont{Spruit}},
  \bibinfo{journal}{A\&A} \textbf{\bibinfo{volume}{450}}, \bibinfo{pages}{1097}
  (\bibinfo{year}{2006}).

\bibitem[{\citenamefont{Braithwaite}(2007)}]{braithwaite07}
\bibinfo{author}{\bibfnamefont{J.}~\bibnamefont{Braithwaite}},
  \bibinfo{journal}{A\&A} \textbf{\bibinfo{volume}{469}}, \bibinfo{pages}{275}
  (\bibinfo{year}{2007}).

\bibitem[{\citenamefont{Braithwaite}(2008)}]{braithwaite08}
\bibinfo{author}{\bibfnamefont{J.}~\bibnamefont{Braithwaite}},
  \bibinfo{journal}{Mon. Not. R. Astron. Soc.} \textbf{\bibinfo{volume}{386}},
  \bibinfo{pages}{1947} (\bibinfo{year}{2008}).

\bibitem[{\citenamefont{Braithwaite}(2009)}]{braithwaite09}
\bibinfo{author}{\bibfnamefont{J.}~\bibnamefont{Braithwaite}},
  \bibinfo{journal}{Mon. Not. R. Astron. Soc.} \textbf{\bibinfo{volume}{397}},
  \bibinfo{pages}{763} (\bibinfo{year}{2009}).

\bibitem[{\citenamefont{Kiuchi et~al.}(2011)\citenamefont{Kiuchi, Yoshida, and
  Shibata}}]{kiuchi11}
\bibinfo{author}{\bibfnamefont{K.}~\bibnamefont{Kiuchi}},
  \bibinfo{author}{\bibfnamefont{S.}~\bibnamefont{Yoshida}}, \bibnamefont{and}
  \bibinfo{author}{\bibfnamefont{M.}~\bibnamefont{Shibata}},
  \bibinfo{journal}{A\&A} \textbf{\bibinfo{volume}{532}}, \bibinfo{pages}{17}
  (\bibinfo{year}{2011}).

\bibitem[{\citenamefont{Lasky et~al.}(2011)\citenamefont{Lasky, Zink, Kokkotas,
  and Glampedakis}}]{lasky11}
\bibinfo{author}{\bibfnamefont{P.~D.} \bibnamefont{Lasky}},
  \bibinfo{author}{\bibfnamefont{B.}~\bibnamefont{Zink}},
  \bibinfo{author}{\bibfnamefont{K.~D.} \bibnamefont{Kokkotas}},
  \bibnamefont{and}
  \bibinfo{author}{\bibfnamefont{K.}~\bibnamefont{Glampedakis}},
  \bibinfo{journal}{Astrophys. J.} \textbf{\bibinfo{volume}{735}},
  \bibinfo{pages}{L20} (\bibinfo{year}{2011}).

\bibitem[{\citenamefont{Ciolfi and Rezzolla}(2012)}]{ciolfi12}
\bibinfo{author}{\bibfnamefont{R.}~\bibnamefont{Ciolfi}} \bibnamefont{and}
  \bibinfo{author}{\bibfnamefont{L.}~\bibnamefont{Rezzolla}},
  \bibinfo{journal}{Astrophys. J.} \textbf{\bibinfo{volume}{760}},
  \bibinfo{pages}{1} (\bibinfo{year}{2012}).

\bibitem[{\citenamefont{Lasky et~al.}(2012)\citenamefont{Lasky, Zink, and
  Kokkotas}}]{lasky12}
\bibinfo{author}{\bibfnamefont{P.~D.} \bibnamefont{Lasky}},
  \bibinfo{author}{\bibfnamefont{B.}~\bibnamefont{Zink}}, \bibnamefont{and}
  \bibinfo{author}{\bibfnamefont{K.~D.} \bibnamefont{Kokkotas}}
  (\bibinfo{year}{2012}), \bibinfo{note}{submitted to Phys. Rev. D,
  arXiv:1203.3590}.

\bibitem[{\citenamefont{Haskell et~al.}(2008)\citenamefont{Haskell, Samuelsson,
  Glampedakis, and Andersson}}]{haskell08}
\bibinfo{author}{\bibfnamefont{B.}~\bibnamefont{Haskell}},
  \bibinfo{author}{\bibfnamefont{L.}~\bibnamefont{Samuelsson}},
  \bibinfo{author}{\bibfnamefont{K.}~\bibnamefont{Glampedakis}},
  \bibnamefont{and}
  \bibinfo{author}{\bibfnamefont{N.}~\bibnamefont{Andersson}},
  \bibinfo{journal}{Mon. Not. R. Astron. Soc.} \textbf{\bibinfo{volume}{385}},
  \bibinfo{pages}{531} (\bibinfo{year}{2008}).

\bibitem[{\citenamefont{Mastrano et~al.}(2011)\citenamefont{Mastrano, Melatos,
  Reissenegger, and Akg\"un}}]{mastrano11}
\bibinfo{author}{\bibfnamefont{A.}~\bibnamefont{Mastrano}},
  \bibinfo{author}{\bibfnamefont{A.}~\bibnamefont{Melatos}},
  \bibinfo{author}{\bibfnamefont{A.}~\bibnamefont{Reissenegger}},
  \bibnamefont{and} \bibinfo{author}{\bibfnamefont{T.}~\bibnamefont{Akg\"un}},
  \bibinfo{journal}{Mon. Not. R. Astron. Soc.} \textbf{\bibinfo{volume}{417}},
  \bibinfo{pages}{2288} (\bibinfo{year}{2011}).

\bibitem[{\citenamefont{Ciolfi et~al.}(2010)\citenamefont{Ciolfi, Ferrari, and
  Gualtieri}}]{ciolfi10}
\bibinfo{author}{\bibfnamefont{R.}~\bibnamefont{Ciolfi}},
  \bibinfo{author}{\bibfnamefont{V.}~\bibnamefont{Ferrari}}, \bibnamefont{and}
  \bibinfo{author}{\bibfnamefont{L.}~\bibnamefont{Gualtieri}},
  \bibinfo{journal}{Mon. Not. R. Astron. Soc.} \textbf{\bibinfo{volume}{406}},
  \bibinfo{pages}{2540} (\bibinfo{year}{2010}).

\bibitem[{\citenamefont{Lander and Jones}(2012)}]{lander12}
\bibinfo{author}{\bibfnamefont{S.~K.} \bibnamefont{Lander}} \bibnamefont{and}
  \bibinfo{author}{\bibfnamefont{D.~I.} \bibnamefont{Jones}},
  \bibinfo{journal}{Mon. Not. R. Astron. Soc.} \textbf{\bibinfo{volume}{424}},
  \bibinfo{pages}{482} (\bibinfo{year}{2012}).

\bibitem[{\citenamefont{{Akg{\"u}n} et~al.}(2013)\citenamefont{{Akg{\"u}n},
  {Reisenegger}, {Mastrano}, and {Marchant}}}]{akgun13}
\bibinfo{author}{\bibfnamefont{T.}~\bibnamefont{{Akg{\"u}n}}},
  \bibinfo{author}{\bibfnamefont{A.}~\bibnamefont{{Reisenegger}}},
  \bibinfo{author}{\bibfnamefont{A.}~\bibnamefont{{Mastrano}}},
  \bibnamefont{and}
  \bibinfo{author}{\bibfnamefont{P.}~\bibnamefont{{Marchant}}},
  \bibinfo{journal}{Mon. Not. R. Astron. Soc.} \textbf{\bibinfo{volume}{433}},
  \bibinfo{pages}{2445} (\bibinfo{year}{2013}), \eprint{1302.0273}.

\bibitem[{\citenamefont{{Ciolfi} and {Rezzolla}}(2013)}]{ciolfi13}
\bibinfo{author}{\bibfnamefont{R.}~\bibnamefont{{Ciolfi}}} \bibnamefont{and}
  \bibinfo{author}{\bibfnamefont{L.}~\bibnamefont{{Rezzolla}}},
  \bibinfo{journal}{Mon. Not. R. Astron. Soc.}  (\bibinfo{year}{2013}),
  \eprint{1306.2803}.

\bibitem[{\citenamefont{Wheeler et~al.}(200)\citenamefont{Wheeler, Yi,
  H\"oflich, and Wang}}]{wheeler00}
\bibinfo{author}{\bibfnamefont{J.~C.} \bibnamefont{Wheeler}},
  \bibinfo{author}{\bibfnamefont{I.}~\bibnamefont{Yi}},
  \bibinfo{author}{\bibfnamefont{P.}~\bibnamefont{H\"oflich}},
  \bibnamefont{and} \bibinfo{author}{\bibfnamefont{L.}~\bibnamefont{Wang}},
  \bibinfo{journal}{Astrophys. J.} \textbf{\bibinfo{volume}{537}},
  \bibinfo{pages}{810} (\bibinfo{year}{200}).

\bibitem[{\citenamefont{Wheeler et~al.}(2002)\citenamefont{Wheeler, Meier, and
  Wilson}}]{wheeler02}
\bibinfo{author}{\bibfnamefont{J.~C.} \bibnamefont{Wheeler}},
  \bibinfo{author}{\bibfnamefont{D.~L.} \bibnamefont{Meier}}, \bibnamefont{and}
  \bibinfo{author}{\bibfnamefont{J.~R.} \bibnamefont{Wilson}},
  \bibinfo{journal}{Astrophys. J.} \textbf{\bibinfo{volume}{568}},
  \bibinfo{pages}{807} (\bibinfo{year}{2002}).

\bibitem[{\citenamefont{Rezzolla et~al.}(2000)\citenamefont{Rezzolla, Lamb, and
  Shapiro}}]{rezzolla00}
\bibinfo{author}{\bibfnamefont{L.}~\bibnamefont{Rezzolla}},
  \bibinfo{author}{\bibfnamefont{F.~K.} \bibnamefont{Lamb}}, \bibnamefont{and}
  \bibinfo{author}{\bibfnamefont{S.~L.} \bibnamefont{Shapiro}},
  \bibinfo{journal}{Astrophys. J.} \textbf{\bibinfo{volume}{531}},
  \bibinfo{pages}{L139} (\bibinfo{year}{2000}).

\bibitem[{\citenamefont{Rezzolla
  et~al.}(2001{\natexlab{a}})\citenamefont{Rezzolla, Lamb, Markovi\'c, and
  Shapiro}}]{rezzolla01a}
\bibinfo{author}{\bibfnamefont{L.}~\bibnamefont{Rezzolla}},
  \bibinfo{author}{\bibfnamefont{F.~K.} \bibnamefont{Lamb}},
  \bibinfo{author}{\bibfnamefont{D.}~\bibnamefont{Markovi\'c}},
  \bibnamefont{and} \bibinfo{author}{\bibfnamefont{S.~L.}
  \bibnamefont{Shapiro}}, \bibinfo{journal}{Phys. Rev. D}
  \textbf{\bibinfo{volume}{64}}, \bibinfo{pages}{104013}
  (\bibinfo{year}{2001}{\natexlab{a}}).

\bibitem[{\citenamefont{Rezzolla
  et~al.}(2001{\natexlab{b}})\citenamefont{Rezzolla, Lamb, Markovi\'c, and
  Shapiro}}]{rezzolla01b}
\bibinfo{author}{\bibfnamefont{L.}~\bibnamefont{Rezzolla}},
  \bibinfo{author}{\bibfnamefont{F.~K.} \bibnamefont{Lamb}},
  \bibinfo{author}{\bibfnamefont{D.}~\bibnamefont{Markovi\'c}},
  \bibnamefont{and} \bibinfo{author}{\bibfnamefont{S.~L.}
  \bibnamefont{Shapiro}}, \bibinfo{journal}{Phys. Rev. D}
  \textbf{\bibinfo{volume}{64}}, \bibinfo{pages}{104014}
  (\bibinfo{year}{2001}{\natexlab{b}}).

\bibitem[{\citenamefont{Cuofano et~al.}(2012)\citenamefont{Cuofano, Dall'Osso,
  Drago, and Stella}}]{cuofano12a}
\bibinfo{author}{\bibfnamefont{C.}~\bibnamefont{Cuofano}},
  \bibinfo{author}{\bibfnamefont{S.}~\bibnamefont{Dall'Osso}},
  \bibinfo{author}{\bibfnamefont{A.}~\bibnamefont{Drago}}, \bibnamefont{and}
  \bibinfo{author}{\bibfnamefont{L.}~\bibnamefont{Stella}},
  \bibinfo{journal}{Phys. Rev. D} \textbf{\bibinfo{volume}{86}},
  \bibinfo{pages}{044004} (\bibinfo{year}{2012}).

\bibitem[{\citenamefont{Ciolfi et~al.}(2011)\citenamefont{Ciolfi, Lander,
  Manca, and Rezzolla}}]{ciolfi11}
\bibinfo{author}{\bibfnamefont{R.}~\bibnamefont{Ciolfi}},
  \bibinfo{author}{\bibfnamefont{S.~K.} \bibnamefont{Lander}},
  \bibinfo{author}{\bibfnamefont{G.~M.} \bibnamefont{Manca}}, \bibnamefont{and}
  \bibinfo{author}{\bibfnamefont{L.}~\bibnamefont{Rezzolla}},
  \bibinfo{journal}{Astrophys. J.} \textbf{\bibinfo{volume}{736}},
  \bibinfo{pages}{L6} (\bibinfo{year}{2011}), \bibinfo{note}{arXiv:1105.3971}.

\bibitem[{\citenamefont{Zimmerman}(1980)}]{zimmerman80}
\bibinfo{author}{\bibfnamefont{M.}~\bibnamefont{Zimmerman}},
  \bibinfo{journal}{Phys. Rev. D} \textbf{\bibinfo{volume}{21}},
  \bibinfo{pages}{891} (\bibinfo{year}{1980}).

\bibitem[{\citenamefont{Zink}(2011)}]{zink11}
\bibinfo{author}{\bibfnamefont{B.}~\bibnamefont{Zink}} (\bibinfo{year}{2011}),
  \bibinfo{note}{arXiv:1102.5202}.

\bibitem[{\citenamefont{Zink et~al.}(2012)\citenamefont{Zink, Lasky, and
  Kokkotas}}]{zink12}
\bibinfo{author}{\bibfnamefont{B.}~\bibnamefont{Zink}},
  \bibinfo{author}{\bibfnamefont{P.~D.} \bibnamefont{Lasky}}, \bibnamefont{and}
  \bibinfo{author}{\bibfnamefont{K.~D.} \bibnamefont{Kokkotas}},
  \bibinfo{journal}{Phys. Rev. D} \textbf{\bibinfo{volume}{85}},
  \bibinfo{pages}{024030} (\bibinfo{year}{2012}).

\bibitem[{\citenamefont{{Yoshida}}(2013)}]{yoshida13}
\bibinfo{author}{\bibfnamefont{S.}~\bibnamefont{{Yoshida}}},
  \bibinfo{journal}{Mon. Not. R. Astron. Soc.}  (\bibinfo{year}{2013}),
  \eprint{1308.1467}.

\bibitem[{\citenamefont{{Mastrano} et~al.}(2013)\citenamefont{{Mastrano},
  {Lasky}, and {Melatos}}}]{mastrano13}
\bibinfo{author}{\bibfnamefont{A.}~\bibnamefont{{Mastrano}}},
  \bibinfo{author}{\bibfnamefont{P.~D.} \bibnamefont{{Lasky}}},
  \bibnamefont{and}
  \bibinfo{author}{\bibfnamefont{A.}~\bibnamefont{{Melatos}}},
  \bibinfo{journal}{Mon. Not. R. Astron. Soc.} \textbf{\bibinfo{volume}{434}},
  \bibinfo{pages}{1658} (\bibinfo{year}{2013}), \eprint{1306.4503}.

\bibitem[{\citenamefont{Zimmerman and Szedenits}(1979)}]{zimmerman79}
\bibinfo{author}{\bibfnamefont{M.}~\bibnamefont{Zimmerman}} \bibnamefont{and}
  \bibinfo{author}{\bibfnamefont{E.}~\bibnamefont{Szedenits}},
  \bibinfo{journal}{Phys. Rev. D} \textbf{\bibinfo{volume}{20}},
  \bibinfo{pages}{351} (\bibinfo{year}{1979}).

\bibitem[{\citenamefont{{van den Broeck}}(2005)}]{vandenbroeck05}
\bibinfo{author}{\bibfnamefont{C.}~\bibnamefont{{van den Broeck}}},
  \bibinfo{journal}{Class. Quantum Grav.} \textbf{\bibinfo{volume}{22}},
  \bibinfo{pages}{1825} (\bibinfo{year}{2005}).

\bibitem[{\citenamefont{Stergioulas and Friedman}(1995)}]{stergioulas95}
\bibinfo{author}{\bibfnamefont{N.}~\bibnamefont{Stergioulas}} \bibnamefont{and}
  \bibinfo{author}{\bibfnamefont{J.~L.} \bibnamefont{Friedman}},
  \bibinfo{journal}{Astrophys. J.} \textbf{\bibinfo{volume}{444}},
  \bibinfo{pages}{306} (\bibinfo{year}{1995}).

\bibitem[{\citenamefont{Stergioulas et~al.}(2004)\citenamefont{Stergioulas,
  Apostolatos, and Font}}]{stergioulas04}
\bibinfo{author}{\bibfnamefont{N.}~\bibnamefont{Stergioulas}},
  \bibinfo{author}{\bibfnamefont{T.~A.} \bibnamefont{Apostolatos}},
  \bibnamefont{and} \bibinfo{author}{\bibfnamefont{J.~A.} \bibnamefont{Font}},
  \bibinfo{journal}{Mon. Not. R. Astron. Soc.} \textbf{\bibinfo{volume}{352}},
  \bibinfo{pages}{1089} (\bibinfo{year}{2004}).

\bibitem[{\citenamefont{Komatsu
  et~al.}(1989{\natexlab{a}})\citenamefont{Komatsu, Eriguchi, and
  Hachisu}}]{komatsu89a}
\bibinfo{author}{\bibfnamefont{H.}~\bibnamefont{Komatsu}},
  \bibinfo{author}{\bibfnamefont{Y.}~\bibnamefont{Eriguchi}}, \bibnamefont{and}
  \bibinfo{author}{\bibfnamefont{I.}~\bibnamefont{Hachisu}},
  \bibinfo{journal}{Mon. Not. R. Astron. Soc.} \textbf{\bibinfo{volume}{237}},
  \bibinfo{pages}{355} (\bibinfo{year}{1989}{\natexlab{a}}).

\bibitem[{\citenamefont{Komatsu
  et~al.}(1989{\natexlab{b}})\citenamefont{Komatsu, Eriguchi, and
  Hachisu}}]{komatsu89b}
\bibinfo{author}{\bibfnamefont{H.}~\bibnamefont{Komatsu}},
  \bibinfo{author}{\bibfnamefont{Y.}~\bibnamefont{Eriguchi}}, \bibnamefont{and}
  \bibinfo{author}{\bibfnamefont{I.}~\bibnamefont{Hachisu}},
  \bibinfo{journal}{Mon. Not. R. Astron. Soc.} \textbf{\bibinfo{volume}{239}},
  \bibinfo{pages}{153} (\bibinfo{year}{1989}{\natexlab{b}}).

\bibitem[{\citenamefont{Ciolfi et~al.}(2009)\citenamefont{Ciolfi, Ferrari,
  Gualtieri, and Pons}}]{ciolfi09}
\bibinfo{author}{\bibfnamefont{R.}~\bibnamefont{Ciolfi}},
  \bibinfo{author}{\bibfnamefont{V.}~\bibnamefont{Ferrari}},
  \bibinfo{author}{\bibfnamefont{L.}~\bibnamefont{Gualtieri}},
  \bibnamefont{and} \bibinfo{author}{\bibfnamefont{J.~A.} \bibnamefont{Pons}},
  \bibinfo{journal}{Mon. Not. R. Astron. Soc.} \textbf{\bibinfo{volume}{397}},
  \bibinfo{pages}{913} (\bibinfo{year}{2009}).

\bibitem[{\citenamefont{Vigelius and Melatos}(2008)}]{vigelius08}
\bibinfo{author}{\bibfnamefont{M.}~\bibnamefont{Vigelius}} \bibnamefont{and}
  \bibinfo{author}{\bibfnamefont{A.}~\bibnamefont{Melatos}},
  \bibinfo{journal}{Mon. Not. R. Astron. Soc.} \textbf{\bibinfo{volume}{386}},
  \bibinfo{pages}{1294} (\bibinfo{year}{2008}).

\bibitem[{\citenamefont{Sathyaprakash and Schutz}(2009)}]{sathyaprakash09}
\bibinfo{author}{\bibfnamefont{B.~S.} \bibnamefont{Sathyaprakash}}
  \bibnamefont{and} \bibinfo{author}{\bibfnamefont{B.~F.}
  \bibnamefont{Schutz}}, \bibinfo{journal}{Living Rev. Relativity}
  \textbf{\bibinfo{volume}{12}}, \bibinfo{pages}{2} (\bibinfo{year}{2009}).

\bibitem[{\citenamefont{Ott et~al.}(2006)\citenamefont{Ott, Burrows, Thompson,
  Livne, and Walder}}]{ott06}
\bibinfo{author}{\bibfnamefont{C.~D.} \bibnamefont{Ott}},
  \bibinfo{author}{\bibfnamefont{A.}~\bibnamefont{Burrows}},
  \bibinfo{author}{\bibfnamefont{T.~A.} \bibnamefont{Thompson}},
  \bibinfo{author}{\bibfnamefont{E.}~\bibnamefont{Livne}}, \bibnamefont{and}
  \bibinfo{author}{\bibfnamefont{R.}~\bibnamefont{Walder}},
  \bibinfo{journal}{Astrophys. J. S.} \textbf{\bibinfo{volume}{164}},
  \bibinfo{pages}{130} (\bibinfo{year}{2006}).

\bibitem[{\citenamefont{Giacomazzo and Perna}(2013)}]{giacomazzo13}
\bibinfo{author}{\bibfnamefont{B.}~\bibnamefont{Giacomazzo}} \bibnamefont{and}
  \bibinfo{author}{\bibfnamefont{R.}~\bibnamefont{Perna}},
  \bibinfo{journal}{Astrophys. J.} \textbf{\bibinfo{volume}{771}},
  \bibinfo{pages}{L26} (\bibinfo{year}{2013}).

\bibitem[{\citenamefont{Glampedakis
  et~al.}(2012{\natexlab{b}})\citenamefont{Glampedakis, Andersson, and
  Lander}}]{glampedakis12a}
\bibinfo{author}{\bibfnamefont{K.}~\bibnamefont{Glampedakis}},
  \bibinfo{author}{\bibfnamefont{N.}~\bibnamefont{Andersson}},
  \bibnamefont{and} \bibinfo{author}{\bibfnamefont{S.~K.}
  \bibnamefont{Lander}}, \bibinfo{journal}{Mon. Not. R. Astron. Soc.}
  \textbf{\bibinfo{volume}{420}}, \bibinfo{pages}{1263}
  (\bibinfo{year}{2012}{\natexlab{b}}).

\bibitem[{\citenamefont{Lander}(2013)}]{lander12b}
\bibinfo{author}{\bibfnamefont{S.~K.} \bibnamefont{Lander}},
  \bibinfo{journal}{Phys. Rev. Lett.} \textbf{\bibinfo{volume}{110}},
  \bibinfo{pages}{071101} (\bibinfo{year}{2013}).

\bibitem[{\citenamefont{Lander et~al.}(2012)\citenamefont{Lander, Andersson,
  and Glampedakis}}]{lander12c}
\bibinfo{author}{\bibfnamefont{S.~K.} \bibnamefont{Lander}},
  \bibinfo{author}{\bibfnamefont{N.}~\bibnamefont{Andersson}},
  \bibnamefont{and}
  \bibinfo{author}{\bibfnamefont{K.}~\bibnamefont{Glampedakis}},
  \bibinfo{journal}{Mon. Not. R. Astron. Soc.} \textbf{\bibinfo{volume}{419}},
  \bibinfo{pages}{732} (\bibinfo{year}{2012}).

\bibitem[{\citenamefont{Mastrano and Melatos}(2012)}]{mastrano12}
\bibinfo{author}{\bibfnamefont{A.}~\bibnamefont{Mastrano}} \bibnamefont{and}
  \bibinfo{author}{\bibfnamefont{A.}~\bibnamefont{Melatos}},
  \bibinfo{journal}{Mon. Not. R. Astron. Soc.} \textbf{\bibinfo{volume}{421}},
  \bibinfo{pages}{760} (\bibinfo{year}{2012}).

\bibitem[{\citenamefont{Pons et~al.}(2009)\citenamefont{Pons, Miralles, and
  Geppert}}]{pons09}
\bibinfo{author}{\bibfnamefont{J.~A.} \bibnamefont{Pons}},
  \bibinfo{author}{\bibfnamefont{J.~A.} \bibnamefont{Miralles}},
  \bibnamefont{and} \bibinfo{author}{\bibfnamefont{U.}~\bibnamefont{Geppert}},
  \bibinfo{journal}{A\&A} \textbf{\bibinfo{volume}{496}}, \bibinfo{pages}{207}
  (\bibinfo{year}{2009}).

\bibitem[{\citenamefont{Gabler et~al.}(2011)\citenamefont{Gabler,
  Cerd\'a-Dur\'an, Font, M\"uller, and Stergioulas}}]{gabler11}
\bibinfo{author}{\bibfnamefont{M.}~\bibnamefont{Gabler}},
  \bibinfo{author}{\bibfnamefont{P.}~\bibnamefont{Cerd\'a-Dur\'an}},
  \bibinfo{author}{\bibfnamefont{J.~A.} \bibnamefont{Font}},
  \bibinfo{author}{\bibfnamefont{E.}~\bibnamefont{M\"uller}}, \bibnamefont{and}
  \bibinfo{author}{\bibfnamefont{N.}~\bibnamefont{Stergioulas}},
  \bibinfo{journal}{Mon. Not. R. Astron. Soc.} \textbf{\bibinfo{volume}{410}},
  \bibinfo{pages}{L37} (\bibinfo{year}{2011}).

\bibitem[{\citenamefont{Gabler et~al.}(2012)\citenamefont{Gabler,
  Cerd\'a-Dur\'an, Stergioulas, Font, and M\"uller}}]{gabler12}
\bibinfo{author}{\bibfnamefont{M.}~\bibnamefont{Gabler}},
  \bibinfo{author}{\bibfnamefont{P.}~\bibnamefont{Cerd\'a-Dur\'an}},
  \bibinfo{author}{\bibfnamefont{N.}~\bibnamefont{Stergioulas}},
  \bibinfo{author}{\bibfnamefont{J.~A.} \bibnamefont{Font}}, \bibnamefont{and}
  \bibinfo{author}{\bibfnamefont{E.}~\bibnamefont{M\"uller}},
  \bibinfo{journal}{Mon. Not. R. Astron. Soc.} \textbf{\bibinfo{volume}{421}},
  \bibinfo{pages}{2054} (\bibinfo{year}{2012}).

\bibitem[{\citenamefont{Vigan\'o and Pons}(2012)}]{vigano12}
\bibinfo{author}{\bibfnamefont{D.}~\bibnamefont{Vigan\'o}} \bibnamefont{and}
  \bibinfo{author}{\bibfnamefont{J.~A.} \bibnamefont{Pons}},
  \bibinfo{journal}{Mon. Not. R. Astron. Soc.} \textbf{\bibinfo{volume}{425}},
  \bibinfo{pages}{2487} (\bibinfo{year}{2012}).

\bibitem[{\citenamefont{Vigan\'o et~al.}(2013)\citenamefont{Vigan\'o, Rea,
  Pons, Perna, Aguilera, and Miralles}}]{vigano13}
\bibinfo{author}{\bibfnamefont{D.}~\bibnamefont{Vigan\'o}},
  \bibinfo{author}{\bibfnamefont{N.}~\bibnamefont{Rea}},
  \bibinfo{author}{\bibfnamefont{J.~A.} \bibnamefont{Pons}},
  \bibinfo{author}{\bibfnamefont{R.}~\bibnamefont{Perna}},
  \bibinfo{author}{\bibfnamefont{D.~N.} \bibnamefont{Aguilera}},
  \bibnamefont{and} \bibinfo{author}{\bibfnamefont{J.~A.}
  \bibnamefont{Miralles}} (\bibinfo{year}{2013}),
  \bibinfo{note}{arXiv:1306.2156}.

\bibitem[{\citenamefont{Gabler et~al.}(2013)\citenamefont{Gabler,
  Cerd\'a-Dur\'an, Font, M\"uller, and Stergioulas}}]{gabler13b}
\bibinfo{author}{\bibfnamefont{M.}~\bibnamefont{Gabler}},
  \bibinfo{author}{\bibfnamefont{P.}~\bibnamefont{Cerd\'a-Dur\'an}},
  \bibinfo{author}{\bibfnamefont{J.~A.} \bibnamefont{Font}},
  \bibinfo{author}{\bibfnamefont{E.}~\bibnamefont{M\"uller}}, \bibnamefont{and}
  \bibinfo{author}{\bibfnamefont{N.}~\bibnamefont{Stergioulas}},
  \bibinfo{journal}{Mon. Not. R. Astron. Soc.} \textbf{\bibinfo{volume}{430}},
  \bibinfo{pages}{1811} (\bibinfo{year}{2013}).

\bibitem[{\citenamefont{Stairs et~al.}(2000)\citenamefont{Stairs, Lyne, and
  Shemar}}]{stairs00}
\bibinfo{author}{\bibfnamefont{I.~H.} \bibnamefont{Stairs}},
  \bibinfo{author}{\bibfnamefont{A.~G.} \bibnamefont{Lyne}}, \bibnamefont{and}
  \bibinfo{author}{\bibfnamefont{S.~L.} \bibnamefont{Shemar}},
  \bibinfo{journal}{Nature} \textbf{\bibinfo{volume}{406}},
  \bibinfo{pages}{484} (\bibinfo{year}{2000}).

\bibitem[{\citenamefont{Durant et~al.}(2013)\citenamefont{Durant, Kargaltsev,
  Pavlov, Kropotina, and Levenfish}}]{durant13}
\bibinfo{author}{\bibfnamefont{M.}~\bibnamefont{Durant}},
  \bibinfo{author}{\bibfnamefont{O.}~\bibnamefont{Kargaltsev}},
  \bibinfo{author}{\bibfnamefont{G.~G.} \bibnamefont{Pavlov}},
  \bibinfo{author}{\bibfnamefont{J.}~\bibnamefont{Kropotina}},
  \bibnamefont{and}
  \bibinfo{author}{\bibfnamefont{K.}~\bibnamefont{Levenfish}},
  \bibinfo{journal}{Astrophys. J.} \textbf{\bibinfo{volume}{763}},
  \bibinfo{pages}{5} (\bibinfo{year}{2013}).

\bibitem[{\citenamefont{Jones and Andersson}(2001)}]{jones01}
\bibinfo{author}{\bibfnamefont{D.~I.} \bibnamefont{Jones}} \bibnamefont{and}
  \bibinfo{author}{\bibfnamefont{N.}~\bibnamefont{Andersson}},
  \bibinfo{journal}{Mon. Not. R. Astron. Soc.} \textbf{\bibinfo{volume}{324}},
  \bibinfo{pages}{811} (\bibinfo{year}{2001}).

\bibitem[{\citenamefont{Levin and {D'Angelo}}(2004)}]{levin04}
\bibinfo{author}{\bibfnamefont{Y.}~\bibnamefont{Levin}} \bibnamefont{and}
  \bibinfo{author}{\bibfnamefont{C.}~\bibnamefont{{D'Angelo}}},
  \bibinfo{journal}{Astrophys. J.} \textbf{\bibinfo{volume}{613}},
  \bibinfo{pages}{1157} (\bibinfo{year}{2004}).

\bibitem[{\citenamefont{Melatos}(2012)}]{melatos12}
\bibinfo{author}{\bibfnamefont{A.}~\bibnamefont{Melatos}},
  \bibinfo{journal}{Astrophys. J.} \textbf{\bibinfo{volume}{761}},
  \bibinfo{pages}{32} (\bibinfo{year}{2012}).

\end{thebibliography}

\end{document}